%% file: main.tex
\PassOptionsToPackage{table}{xcolor}
\documentclass[sigconf]{acmart}
\AtBeginDocument{%
  }

 \setcopyright{none}
  \acmConference[CCS '26]{ACM Conference on Computer and Communications Security (CCS)}{November 15--19 2026}{The Hague, Netherlands}

\usepackage{comment}
\usepackage{xspace}
\usepackage{graphicx}
\usepackage{makecell}
\usepackage{adjustbox}
\usepackage{tikz}
\usepackage{multicol}
\usepackage{multirow}
\usepackage{amsmath}
\usepackage{mathpartir}
\usepackage{pifont}
\usepackage{algorithm}
\usepackage{algorithmic}
\usepackage{booktabs}

\usepackage{amssymb}
\usepackage{dashbox}
\usepackage{subcaption}
\usepackage{threeparttable}
\usepackage{placeins}
\usepackage{url}
\usepackage{makecell} 
\usepackage{tcolorbox}
\tcbuselibrary{listings,skins}
\usepackage{cleveref}
\usepackage{wasysym}
\usepackage[table]{xcolor}   
\definecolor{SOne}{RGB}{255,230,230}  
\definecolor{STwo}{RGB}{230,240,255}  
\usepackage{multirow}        
\usepackage{array}
\usepackage{listings}
\usetikzlibrary{shapes.geometric, arrows, positioning}
\usepackage{soul}

\usepackage{caption}

\newfloat{listing}{tbp}{lol}
\floatname{listing}{Listing}

\definecolor{codegreen}{rgb}{0,0.6,0}
\definecolor{codegray}{rgb}{0.5,0.5,0.5}
\definecolor{codepurple}{rgb}{0.58,0,0.82}
\definecolor{backcolour}{rgb}{0.95,0.95,0.92}
\definecolor{darkgreen}{RGB}{0,128,0}
\definecolor{deepskyblue}{RGB}{0,150,255}
\definecolor{amber}{RGB}{255,130,0}
\definecolor{red}{RGB}{200,0,0}
\definecolor{blue}{RGB}{0,0,180}

\newcommand{\legenditem}[2]{%
  \textcolor{#1}{\rule{6pt}{6pt}}\,#2%
  }

\definecolor{basicTEE}{HTML}{1F77B4}   
\definecolor{teeIFT}{HTML}{FF7F0E}     
\definecolor{teeIFTDec}{HTML}{2CA02C}  

\tikzstyle{node} = [circle, draw, text centered, minimum size=0.8cm]
\tikzstyle{arrow} = [thick,->,>=stealth]

\newcommand{\kingsguard}{{\tt KINGS\-GUARD}\xspace}
\newcommand{\kgrule}[3]{\mprset{vskip=1ex}\inferrule[\noindent\fbox{\parbox{5cm}{\centering \textsc{#1}}}]{#2}{#3}}

\usepackage{fancyhdr}
\fancypagestyle{firstpage}{%
  \fancyhf{}
  \fancyhead[C]{%
    \begin{tcolorbox}[
      colframe=black,
      boxrule=0.3pt,
      arc=0pt,
      left=4pt,right=4pt,top=2pt,bottom=2pt,
      width=\textwidth
    ]
    \footnotesize\normalfont
    If you cite this paper, please use the CCS reference: \textit{Saltanat 
    Firdous Allaqband, Deepanjali S, Rohit Srinivas R G, Devashish Gosain, 
    and Chester Rebeiro. 2026. KingsGuard: Enclave Data Protection Under 
    Real-World TEE Vulnerabilities. In Proceedings of the 2026 ACM SIGSAC 
    Conference on Computer and Communications Security (CCS '26), November 
    15--19, 2026, The Hague, Netherlands. ACM, New York, NY, USA.}
    \end{tcolorbox}%
  }
  
}

\renewcommand\footnotetextcopyrightpermission[1]{}
\begin{document}

\title{KingsGuard: Enclave Data Protection Under Real-World TEE Vulnerabilities}

\author{Saltanat Firdous Allaqband}
\email{saltanat@cse.iitm.ac.in}
\affiliation{%
  \institution{Indian Institute of Technology Madras}
  \country{India}
}
\author{Deepanjali S}
\email{ic39819@imail.iitm.ac.in}
\affiliation{%
  \institution{Indian Institute of Technology Madras}
  \country{India}
}

\author{Rohit Srinivas R G}
\email{cs23s046@smail.iitm.ac.in}
\affiliation{%
  \institution{Indian Institute of Technology Madras}
  \country{India}
}

\author{Devashish Gosain}
\email{dgosain@cse.iitb.ac.in}
\affiliation{%
  \institution{Indian Institute of Technology Bombay}
  \country{India}
}

\author{Chester Rebeiro}
\email{chester@cse.iitm.ac.in}
\affiliation{%
  \institution{Indian Institute of Technology Madras}
  \country{India}
}

\renewcommand{\shortauthors}{Saltanat et al.}

\begin{abstract}
Trusted Execution Environments (TEEs) have emerged as a cornerstone for securing sensitive computations by providing isolated enclaves protected from untrusted software. However, their security guarantees are undermined by vulnerabilities in both the enclave code and the underlying hardware design, which can allow sensitive data to leak despite strong isolation guarantees. This paper presents \kingsguard, a novel TEE design that systematically monitors and controls the propagation of sensitive data within an enclave. By enforcing fine-grained data flow tracking and checks in hardware, our approach ensures that sensitive data does not leave the enclave boundary, thus bridging the gap between the idealized threat models of TEEs and their practical realizations. Additionally, to balance security with practical functionality, we introduce controlled declassification at enclave boundaries, allowing intentional release of data to the outside world.
Our implementation of \kingsguard on a RISC-V processor has a 10.8\% hardware area overhead when synthesized on FPGA and a 5.69\% performance overhead. 
\end{abstract}




\setlength{\headheight}{35pt}
\addtolength{\topmargin}{-23pt}

\maketitle
\pagestyle{plain}
\thispagestyle{firstpage}
\section{Introduction}
\label{sec:intro}
\input{Introduction}

\section{Background}
\label{sec:bg}
\input{Background}
\section{Threat Model and Assumptions}
\label{sec:threat}
\input{threatmodel}

\section{Preventing Data Exfiltration from Enclaves}
\label{sec:overview}
\input{Overview}

\section{\kingsguard Design}
\label{sec:design}

\input{Design}

\section{Implementation}
\label{sec:implementation}
\input{Implementation}

\section{Experimental Setup and Results}
\label{sec:results}
\input{Results}

\section{Security Analysis}
\label{sec:secanal}
\input{Security}

\section{Related Work}
\label{sec:rw}
\input{RelatedWork}
\section{Conclusion}
\label{sec:conclusion}
\input{Conclusion}

{\flushleft \bf Acknowledgement.} This work is funded by the Ministry of Electronics and Information Technology, India (MeitY) and Industrial Development Bank of India (IDBI). We also acknowledge the use of ChatGPT for minor editorial improvements, including grammar and language polishing. 

\section*{Ethical Considerations}
The presented work concerns TEE security research. During this research ,no human subjects were
involved at any point. We do not find any new vulnerabilities or attacks in any system, but replicate already existing attacks to demonstrate the security of our solution. We conduct all our experiments on our in-lab simulation of a RISC-V processor and it raises no ethical considerations. We are confident that our research has not violated any legal standards, is in the interest of computer security around the world and adheres to the USENIX Security ’25 Ethics Guidelines.


\bibliographystyle{ACM-Reference-Format}
\bibliography{biblio}


\end{document}

%% file: Introduction.tex
Trusted Execution Environments (TEEs)~\cite{DBLP:journals/iacr/sgxCostanD16, beniamini2015trustzone} have gained significant popularity in recent years due to their ability to strengthen system security. At their core, TEEs provide hardware mechanisms to isolate regions of execution, called {\em enclaves}. These enclaves shield sensitive code and data from unauthorized access even in the presence of compromised or malicious privileged software, such as the Operating System (OS) and the hypervisor, thus providing robust guarantees of confidentiality and integrity.

However, the security guarantees provided by TEEs rely on a strong underlying assumption that the enclave code and the TEE platform are free from vulnerabilities and flaws. This assumption is unrealistic given that modern hardware and software designs are extremely complex and highly prone to design and implementation flaws. This places a considerable burden on developers to produce code that is not just functionally correct but also free of all vulnerabilities. Consequently, placing unconditional trust in the enclave leads to a fragile security model that overlooks the existence of bugs within the enclave code or design. If such vulnerabilities are discovered, the very foundation of the enclave’s isolation can be undermined, leaving sensitive data exposed.


Vulnerabilities in TEEs may arise from both software bugs in enclave code~\cite{DBLP:conf/uss/BiondoCDFS18, lee2017hacking, yoon2022sgxdump, cloosters2020teerex, khandaker2020coin, weichbrodt2016asyncshock} or hardware-level weaknesses~\cite{cui2021smashex, DBLP:conf/uss/BorrelloKSLG022, cve2016_10423, m1racles2021}.
Software bugs, such as memory safety violations, can be exploited to extract secrets from the enclaves either directly~\cite{lee2017hacking, DBLP:conf/uss/BiondoCDFS18, khandaker2020coin, weichbrodt2016asyncshock} or indirectly~\cite{yoon2022sgxdump}. On the other hand, hardware-level weaknesses arise from shared hardware resources such as caches, registers or peripheral buses. Secrets from the enclave may affect the state of these resources. This may lead to indirect leakage of data via \emph{timing channels}~\cite{van2018foreshadow, DBLP:conf/uss/BorrelloKSLG022, ge2018surveytiming, gotzfried2017cacheattacksonsgx, schwarz2017malware}, where secrets are inferred from execution latency variations (e.g., cache hit/miss behavior), or \emph{storage channels}~\cite{m1racles2021, cve2016_10423, CVE-2017-6296}, where secrets are encoded in the architecturally visible state of shared hardware structures, such as registers, and subsequently observed by an attacker.
Although considerable research has focused on mitigating timing channels  (e.g.,~\cite{DBLP:conf/uss/sanctumCostanLD16, DBLP:conf/ndss/zerotrace, domnitser2012non, sanchez2011vantage, wang2016secdcp, dessouky2020hybcache}), comparatively little attention has been given to preventing sensitive data leaks caused by enclave software bugs or hardware storage channels. 
The consequences of exploiting these bugs are severe, enabling attackers to perform key extraction~\cite{cve2016_2431}, escalate privileges~\cite{suciu2020hpe}, or even obtain complete dumps of enclave memory~\cite{yoon2022sgxdump}.

{\noindent\textbf{Our goal:}} In this paper, we aim to strengthen the isolation guarantees of TEEs by introducing a novel TEE framework, \kingsguard, that prevents leakage of sensitive data from enclaves even in the presence of software bugs and hardware storage channels. Unlike existing approaches~\cite{DBLP:journals/iacr/sgxCostanD16, DBLP:conf/coinco/trustzoneNgabonzizaMBCM16, DBLP:conf/uss/sanctumCostanLD16, DBLP:conf/eurosys/keystoneLeeKSAS20, DBLP:conf/uss/cureBahmaniBDJKSS21, DBLP:conf/uss/sancusNoormanADSHHPVP13, DBLP:conf/ndss/sanctuaryBrasserGJSS19, DBLP:conf/eurosys/trustliteKoeberlSSV14} that assume ``flawless enclaves'', \kingsguard is designed to operate under realistic scenarios where these software bugs and hardware weaknesses may be exploited. \kingsguard thus bridges a critical gap between the idealized assumptions and the practical realities of the TEE security model. It achieves this by actively detecting and preventing unauthorized data flows from enclaves. At a high level, \kingsguard taints and tracks sensitive information during execution as it flows through the enclave and prevents tainted information from escaping the enclave. Designing such a framework, however, introduces the following key challenges:

{\flushleft\textbf{Challenge C1:}} Information flow tracking traditionally requires source-level modifications or OS support to insert taint propagation logic~\cite{tarkhani2023deluminator, DBLP:journals/corr/abs-2401-08901, DBLP:conf/uss/TsaiSJMPP20}. Such requirements make this approach incompatible with existing applications or expand the Trusted Computing Base (TCB) by relying on the OS. Achieving transparency and compatibility, therefore, is a significant challenge. \kingsguard should thus avoid extensive modifications to the application or other software components. 

{\noindent\textbf{Challenge C2:}} Traditional taint tracking is useful for protecting resources that are exclusive to an enclave, such as its private memory. However, certain resources, such as the page tables, registers, I/O peripherals, and interconnect buses, are shared across the enclave and non-enclave boundaries. 
This shared state can inadvertently leak enclave data 
even without explicitly copying sensitive data outside the enclave. \kingsguard should be able to prevent such indirect leakage.

{\noindent\textbf{Challenge C3:}} 
As part of their regular functionality, enclaves often legitimately interact with untrusted software, for example, to return computation results, perform system calls, or communicate with untrusted services.
 A na\"{i}ve taint-tracking mechanism that prevents any tainted information from leaving the enclave would cripple such functionality. To support enclave interaction with the non-enclave code, \kingsguard should therefore be able to distinguish between authorized and unauthorized data flows across the enclave boundary. Authorized flows are incorporated by design to enable communication between enclave and non-enclave regions. These flows should be permitted, while unauthorized data flows should be blocked. 

A design principle that helps \kingsguard address all three challenges is \textbf{\textit{implementing dynamic information flow tracking entirely in hardware}}. To the best of our knowledge, \kingsguard is the first TEE framework to employ fine-grained hardware-enforced information flow tracking to prevent sensitive data leakage from enclaves caused due to bugs in the enclave code and shared hardware storage channels.
The hardware-centric approach results in low performance overheads and also minimizes changes needed to the software. \kingsguard works directly with compiled binaries with no modifications required to the application source code or the OS, other than introducing minor user annotations, thus addressing \textbf{C1} (\S~\ref{sec:bin_prep}). To address \textbf{C2}, \kingsguard extends taint tracking beyond data to addresses and shared hardware registers to prevent indirect data leakage (\S~\ref{sec:dift}). To address \textbf{C3}, \kingsguard extends the hardware to ensure that any data release from the enclave occurs only along \emph{authorized declassification paths}, thus distinguishing between authorized and unauthorized data flows (\S~\ref{sec:declass}).

To demonstrate practical feasibility, we implement \kingsguard on a RISC-V processor and synthesize the design on an FPGA. We further integrate \kingsguard into the gem5 simulator~\cite{DBLP:journals/sigarch/BinkertBBRSBHHKSSSSVHW11}, a cycle-accurate platform widely used in both academia and industry, enabling rapid evaluation of multiple system configurations. While \kingsguard is designed to prevent data leakage in enclaves caused due to software bugs and hardware storage channels, it can also be integrated with existing countermeasures for timing channels~\cite{DBLP:conf/sp/GinerSPEUMG23, DBLP:conf/uss/WernerUG0GM19scattercache, tan2020phantomcache, DBLP:conf/micro/Qureshi18, DBLP:conf/micro/mi6} to provide stronger security guarantees with marginal overheads as demonstrated in Section~\ref{sec:secanal}.  We perform a  security evaluation of \kingsguard, showing its effectiveness against multiple attack vectors that exploit software vulnerabilities and hardware flaws to leak data directly by copying it to non-enclave memory or indirectly via shared resources. We have also shown \kingsguard's effectiveness in preventing data leakage in a Supervisory Control And Data Acquisition~(SCADA) application.



We summarize our contributions as follows.
\begin{enumerate}
    \item We propose \kingsguard, a TEE that protects applications in an enclave despite the presence of exploitable software bugs or hardware design flaws using an Information Flow Tracking mechanism implemented entirely in hardware, while ensuring that legitimate data is released securely through authorized declassification paths.
    \item \kingsguard provides an application-agnostic solution that does not require extensive modifications to the applications, OS, or any other software components, ensuring compatibility and transparency.
    \item \kingsguard prevents data leakage from enclaves, whether direct through explicit copying, or indirect through hardware storage channels. 
    \item We extend a RISC-V processor to implement \kingsguard and evaluate it on an FPGA and a simulator, showing that it incurs an average performance overhead of 5.69\% and hardware area overhead of 10.8\%.
    \item We extend \kingsguard with a state-of-the-art cache-based timing side-channel countermeasure, SassCache~\cite{DBLP:conf/sp/GinerSPEUMG23} to demonstrate its compatibility with existing side-channel defenses.
    \item We provide a security analysis of \kingsguard and demonstrate its effectiveness in mitigating data leakage using a SCADA application.
\end{enumerate}




%% file: Background.tex
\begin{table}[!t]
\centering
\caption{Categorization of data leakage attacks based on source of vulnerability (software or hardware), and exfiltration channel (direct or indirect).
}
\resizebox{\columnwidth}{!}{%
\begin{tabular}{|l|l|l|l|}
\hline
\textbf{Source} & \textbf{Attacks} & \textbf{Channel} & \textbf{Vulnerability} \\ \hline

\multirow{8}{*}{Software} 
& CVE-2015-6639~\cite{beniamini2015android} & Direct & \multirow{5}{*}{\centering \shortstack{Memory Corruption}} \\ \cline{2-3}

 & CVE-2016-2431~\cite{beniamini2015trustzone} & Direct &  \\ \cline{2-3}
 & Pointer-Based Data Leakage~\cite{stella} & Direct &  \\ \cline{2-3}
 & Hacking in Darkness~\cite{lee2017hacking} & Direct &  \\ \cline{2-3}
 & Guards Dilemma \cite{DBLP:conf/uss/BiondoCDFS18}& Direct &  \\ \cline{2-3}
 & SGXDump \cite{yoon2022sgxdump} & Indirect &  \\ \cline{2-4}
 
 & AsyncShock \cite{weichbrodt2016asyncshock} & Direct & \multirow{3}{*}{TOCTOU} \\ \cline{2-3}
 & COIN \cite{khandaker2020coin} & Direct &  \\ \cline{2-3}
 & CVE-2017-6296~\cite{CVE-2017-6296} & Direct &  \\ \hline

\multirow{4}{*}{Hardware} 
& SmashEx ~\cite{cui2021smashex} & Direct & Non-Atomic Exception Handling \\ \cline{2-4}
 & CVE-2016-10423~\cite{cve2016_10423} & Indirect & Non-exclusive shared I/O \\ \cline{2-4}
 & M1racles ~\cite{m1racles2021} & Indirect & Missing Register Access-Control \\ \hline

\end{tabular}%
}
\label{tab:attacks}
\end{table}

Vulnerabilities in TEEs originate from either the software or hardware, potentially enabling leakage of sensitive data from the enclave. The leakage can occur either by direct copying of data to memory regions outside the enclave or indirectly through shared resources.
\Cref{tab:attacks} presents a taxonomy of data leaks reported in the literature, categorized according to these dimensions. 

{\flushleft \bf Sources of Vulnerabilities in TEEs.}
\emph{Software bugs} in enclave code, like buffer overflows~\cite{beniamini2015android,lee2017hacking}, use-after-free~\cite{khandaker2020coin}, improper sanitization~\cite{stella}, and TOCTOU~\cite{weichbrodt2016asyncshock,khandaker2020coin, CVE-2017-6296}, can be exploited to build ROP gadgets that hijack the enclave's control flow to leak sensitive data into non-enclave regions. 
On the other hand, \emph{hardware vulnerabilities} in TEE designs fail to ensure strict isolation~\cite{cve2016_10423} and atomicity~\cite{cui2021smashex} of enclave state, allowing shared hardware resources to be exploited for data leakage. These resources may include registers~\cite{m1racles2021} or I/O interfaces~\cite{cve2016_10423} and can expose sensitive information directly from hardware, even if the enclave code is meticulously designed without software bugs. 

{\noindent \bf Channels of Data Exfiltration from Enclaves.}
Vulnerabilities in TEEs can be exploited to leak sensitive data either through \emph{direct} or \emph{indirect} channels. Direct leakage occurs when sensitive data is explicitly copied from the enclave, e.g., using \texttt{memcpy}~\cite{lee2017hacking, DBLP:conf/uss/BiondoCDFS18, khandaker2020coin, weichbrodt2016asyncshock}. 
Besides direct copying, data can also leak from the enclave indirectly via shared resources~\cite{yoon2022sgxdump, m1racles2021, cve2016_10423}. This indirect leakage can either be timing-based or storage-based. Timing channels exploit variations in resource access times to interpret sensitive information~\cite{van2018foreshadow, gotzfried2017cacheattacksonsgx}.
In contrast, storage channels exploit shared storage resources that are not intended for communication, where improper isolation or observable residual state enables an adversary to exfiltrate information~\cite{cve2016_10423, yoon2022sgxdump, m1racles2021}. For instance, the M1racles bug in Apple M1~\cite{m1racles2021} exploited a shared register in the user space to establish covert communication between two processes. Such a shared register can also be used to leak sensitive data from an enclave.

\kingsguard prevents data leaks originating from both software and hardware vulnerabilities, whether through direct copying or through indirect storage channels. It can prevent all attacks mentioned in \Cref{tab:attacks} and can be extended with existing defenses against timing side channels as shown in Section~\ref{sec:secanal}.

%% file: threatmodel.tex
We assume that the OS and other commodity software running in the system may be compromised, while protected applications run in isolated enclaves. Unlike traditional TEE threat models, we explicitly recognize that the enclave code and the TEE platform may contain exploitable vulnerabilities, and our design accounts for the possibility that an attacker could attempt to leverage such vulnerabilities to exfiltrate sensitive information, either through direct copying or indirectly via storage channels. \kingsguard thus strengthens the threat model beyond the conventional assumption of bug-free enclaves, addressing a more realistic adversary. 

We assume a trusted software layer, called the \emph{Security Monitor} (SM), running at the highest privilege level in the CPU. 
We assume that the CPU supports secure boot, ensuring that each stage of the boot process is authenticated and integrity-verified using signatures anchored in hardware. This guarantees that only authorized, untampered enclave code can run, and that the SM is itself trustworthy, even when the rest of the software stack is compromised.

Certain classes of attacks remain out of scope. In particular, denial-of-service (DoS) attacks by a malicious OS, are not considered. Additionally, physical attacks such as probing, fault injection, or invasive tampering are all out of scope. \kingsguard's design does not prevent timing side-channel attacks. However, it is designed to be compatible with existing countermeasures~\cite{DBLP:conf/uss/sanctumCostanLD16, DBLP:conf/micro/mi6, DBLP:conf/ndss/zerotrace} against these attacks.

%% file: Overview.tex
\kingsguard prevents data exfiltration from enclaves caused due to software bugs or hardware storage channels. It achieves this by integrating hardware-assisted information flow tracking that continuously monitors the flow of sensitive information during enclave execution and permits only authorized data to leave the enclave. 
This section outlines how \kingsguard prevents unauthorized data flows from enclaves, whether through direct copying to non-enclave memory or indirect leakage via shared resources, while still allowing authorized flow of information across the enclave.

\subsection{Preventing Data Leaks by Direct Copying}
\label{sec:over-dir}
\begin{figure}[!t]
    \centering \includegraphics[width=0.45\textwidth]{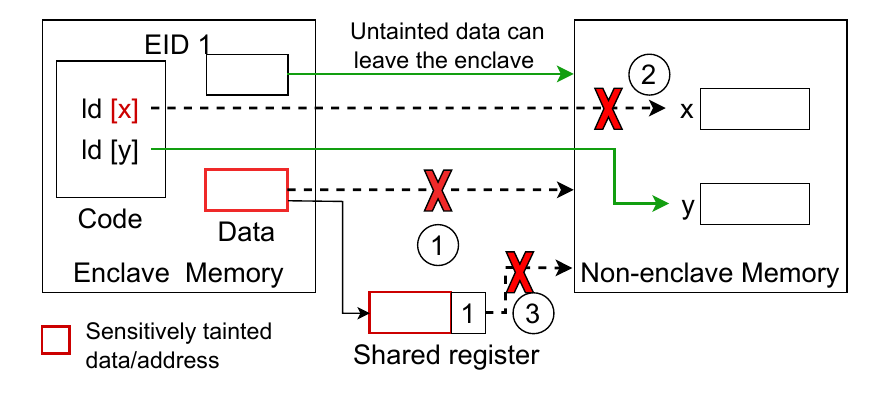}
    \caption{\kingsguard prevents enclave data leakage through three enforcement points: (1) checking taints at the enclave boundary to stop direct leakage into untrusted memory, (2) tracking whether sensitive values influence non-enclave addresses to block indirect leakage, (3) stamping shared registers to prevent unauthorized reads from outside the enclave.
    \label{fig:dift}}
\end{figure}

\kingsguard uses \textit{taint tracking} to prevent unauthorized leakage caused by direct copying of sensitive data from enclave to non-enclave memory. Sensitive data in an enclave is marked with taints post-compilation (see \S~\ref{sec:bin_prep}). During execution, these taints are propagated through all computations and memory operations at the hardware level, ensuring that any values derived from tainted data are also tainted. To prevent tainted data from leaving the enclave boundary, every store operation to non-enclave memory is checked in hardware. If the data is tainted, it is not stored to non-enclave memory (see Figure~\ref{fig:dift}) unless it is explicitly authorized to exit the enclave through a \emph{declassification} mechanism. We discuss more on declassification in Section~\ref{sec:authorized}.

\subsection{Preventing Indirect Data Leaks through Shared Resources}
\label{sec:over-indirect}
In addition to tracking data, \kingsguard monitors the influence of sensitive information on memory addresses. Addresses derived from tainted values can become a channel for leakage by encoding secrets as memory accesses. This attack vector was demonstrated in SGXDump~\cite{yoon2022sgxdump}, where enclave secrets were exfiltrated by encoding them into page tables. 
To mitigate this threat, \kingsguard checks every enclave-generated address pointing to the non-enclave memory and denies access if the address is tainted. For example, in Figure~\ref{fig:dift}, enclave code is trying to access addresses {\tt x} and {\tt y} in the non-enclave memory. {\tt x} is tainted, while {\tt y} is not tainted. \kingsguard allows the access to {\tt y}, but not {\tt x}. By enforcing taint tracking over both data and addresses, \kingsguard blocks a wide spectrum of exfiltration attempts.

Shared registers between enclave and non-enclave modes can also be exploited to leak sensitive information. Data can be placed in the shared hardware resource by enclave code and read by the non-enclave code. Such leakage cannot be detected by enforcing checks on store operations to non-enclave memory, as is done for detecting malicious direct memory copies.

\kingsguard prevents such leakage by extending its tainting mechanism to shared hardware registers with an identifier unique to the enclave, called the \emph{Enclave Identifier} (EID). When an enclave writes tainted data to a shared register, the register is stamped with that enclave’s EID as shown in Figure~\ref{fig:dift}. Subsequent reads from the register are allowed only to the enclave with the same EID. Non-enclave code or an enclave with a different EID is not allowed to read the value from the stamped register. However, a different enclave can overwrite the register and update its EID, thereby \emph{restamping} it. Thus, an enclave is not allowed to read tainted data from a shared register unless it has overwritten it. This mechanism prevents an attacker from reading sensitive enclave data from shared registers.




\subsection{Distinguishing Authorized from Unauthorized Data Flows}\label{sec:authorized}
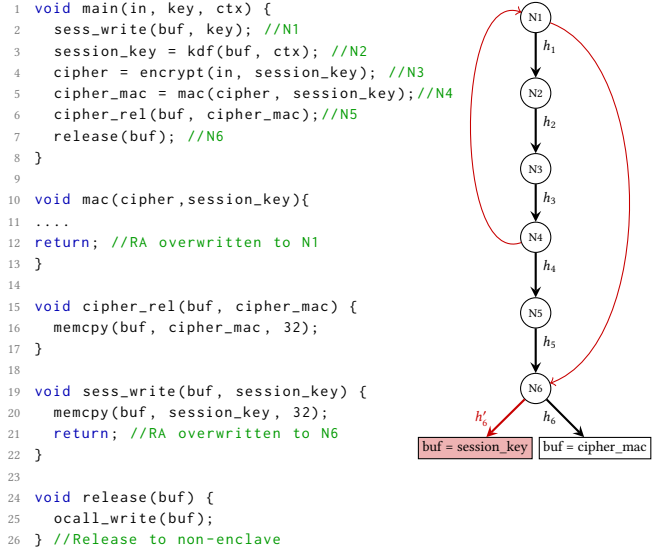
\begin{figure}[!t]
\centering
\begin{minipage}[t]{0.3\columnwidth}

\vspace{2pt}
\begin{lstlisting}
void main(in, key, ctx) {
  sess_write(buf, key); //N1
  session_key = kdf(buf, ctx); //N2
  cipher = encrypt(in, session_key); //N3
  cipher_mac = mac(cipher, session_key);//N4
  cipher_rel(buf, cipher_mac);//N5
  release(buf); //N6
}  

void mac(cipher,session_key){
....
return; //RA overwritten to N1
}

void cipher_rel(buf, cipher_mac) {
  memcpy(buf, cipher_mac, 32);
}

void sess_write(buf, session_key) {
  memcpy(buf, session_key, 32);
  return; //RA overwritten to N6
}  

void release(buf) {
  ocall_write(buf);   
} //Release to non-enclave
\end{lstlisting}
\end{minipage}%
\hfill%
\begin{minipage}[t]{0.40\columnwidth}
\vspace{1pt}
\begin{tikzpicture}[node distance=2.0cm, scale=0.7, every node/.style={scale=0.5}]

\node[node] (N1) {N1};
\node[node, below of=N1] (N2) {N2};
\node[node, below of= N2] (N3) {N3};
\node[node, below =0.5 cm of N3] (N4) {N4};
\node[node, below of=N4] (N5) {N5};
\node[node, below of=N5] (N6) {N6};

\node[below right=0.1cm and -0.1cm of N1, text width=4cm, align=left, font=\Large] {$h_1$};
\node[below right=0.1cm and -0.1cm of N2, text width=3.5cm, align=left, font=\Large] {$h_2$};
\node[below right=0.1cm and -0.1cm of N3, text width=3.5cm, align=left, font=\Large] {$h_3$};
\node[below right=0.1cm and -0.1cm of N4, text width=3.5cm, align=left, font=\Large] {$h_4$};
\node[below right=0.1cm and -0.1cm of N5, text width=3.5cm, align=left, font=\Large] {$h_5$};
\node[below right=0.1cm and -0.1cm of N6, text width=4cm, align=left, font=\Large] {$h_6$};
\node[text=red][below left=0.1cm and -1.4cm of N6, text width=4cm, align=left, font=\Large ]{$h_6'$};


\node[below right=0.5cm and -0.1cm of N6,
      draw, rectangle, align=center, font=\Large] (buf1)
      {buf = cipher\_mac};

\node[below left=0.5cm and -0.1cm of N6,
      draw, rectangle, fill=red!30, align=center, font=\Large] (buf2)
      {buf = session\_key};

\draw[arrow] (N1) -- (N2);
\draw[arrow] (N2) -- (N3);
\draw[arrow] (N3) -- (N4);
\draw[arrow] (N4) -- (N5);
\draw[arrow] (N5) -- (N6);
\draw[->, red] (N1) to[out=-20, in=20, looseness=0.8] (N6);

\draw[->, red] (N4) to[out=200, in=160, looseness=0.8] (N1);

\draw[arrow] (N6) -- (buf1); 
\draw[arrow, red] (N6) -- (buf2); 

\end{tikzpicture}
\end{minipage}
\caption{An enclave code snippet with its corresponding control flow graph showing the ADP~($h6$) for \texttt{cipher\_mac}. The Return Address~(RA), overwritten at lines 12 and 21, hijacks the control flow to an unauthorized path~($h6'$), exposing the \texttt{session\_key}. This violation is detected by \kingsguard.}
\label{fig:declassification}
\end{figure}
While taint tracking effectively confines sensitive data within the enclave memory, legitimate interactions with non-enclave components sometimes require controlled release of such data. To enable these sanctioned flows, \kingsguard employs \textit{declassification}~\cite{sabelfeld2005dimensions}, a mechanism that selectively removes taints from data authorized for release. \kingsguard relies on predefined \emph{Authorized Declassification Paths} (ADPs), which are identified during build time and added to the enclave binary to authorize the release of tainted data from the enclave. We provide more details about ADPs in Section~\ref{sec:bin_prep}. Only data that traverses any of these predefined ADPs is permitted to cross the enclave boundary. At runtime, the hardware maintains a running cumulative hash, updating it upon each control-flow instruction. When data needs to be written out of the enclave boundary, this runtime hash is compared against the set of pre-computed ADPs to verify the legitimacy of the execution path. The release of tainted data is permitted only if the hashes match, indicating that execution followed an authorized path. 


Consider the enclave code snippet in Figure~\ref{fig:declassification}. Each node represents a control flow instruction; for instance, N1 represents the call to function \texttt{sess\_write()}. A control transfer has a source address and a destination address. Source address in this case is the address of $N1$ and destination address is the entry address of \texttt{sess\_write()}. $h1$ is computed from a combination of an initial constant, the address of $N1$, and the entry address of \texttt{sess\_write()}, whereas subsequent hashes ($h2-h6$) use the next instruction’s source and destination addresses along with the previous hash, yielding a cumulative hash. In the legitimate execution path \texttt{N1$\rightarrow$N2$\rightarrow$N3$\rightarrow$N4$\rightarrow$N5$\rightarrow$N6}, the variable \texttt{cipher\_\allowbreak mac} is written to the \texttt{buf} that is released outside the enclave. The control flow of this legitimate path is statically enumerated and represented as an ADP that allows the release of \texttt{cipher\_mac}. However, the alternate path \texttt{N1$\rightarrow$N2$\rightarrow$N3$\rightarrow$N4$\rightarrow$N1$\rightarrow$\allowbreak N6} reflects a hijacked control flow where the \texttt{session\_key} is copied into the \texttt{buf} for release. At $N6$, where \texttt{ocall\_write()} stores data to non-enclave memory, declassification is performed, and the runtime hash is compared to precomputed ADPs. A match ($h6$) indicates a legitimate release, removing the taint from \texttt{cipher\_mac} and allowing it to be released from the enclave. However, on a mismatch ($h6'$), the data being written out (\texttt{session\_key}) is replaced with zeroes, preventing its leakage.  

%% file: Design.tex
\kingsguard's design aims to uphold the strong isolation guarantees expected of TEEs by protecting against unauthorized leakage arising from programming vulnerabilities and hardware storage channels. \kingsguard prevents sensitive information from leaving the enclave, and at the same time enables legitimate outputs to be released in a secure and controlled manner. This section describes the design of \kingsguard to achieve these objectives. 
We first explain how a binary is prepared with minor annotations (\S~\ref{sec:bin_prep}), addressing challenge \textbf{C1}. Next, we describe how \kingsguard preserves the baseline TEE guarantees(\S~\ref{sec:tee}), then explain how it prevents unauthorized data leakage using hardware-supported information flow tracking (\S~\ref{sec:dift}), addressing challenge \textbf{C2}, and finally show how it supports controlled release of legitimate outputs through secure declassification (\S~\ref{sec:declass}), addressing challenge \textbf{C3}. We use formal notations to explain the design of \kingsguard clearly.\footnote{
\[
\centering
\kgrule{rulename}{
\text{Premise-1} \\ 
\text{Premise-2} \ldots \text{Premise-N}
}{\text{Conclusion}}
\]
Change in conclusion occurs only if all premises execute without failing.}

\subsection{Binary Preparation for \kingsguard}\label{sec:bin_prep}
\begin{figure}[!t]
    \includegraphics[width=0.3\textwidth]{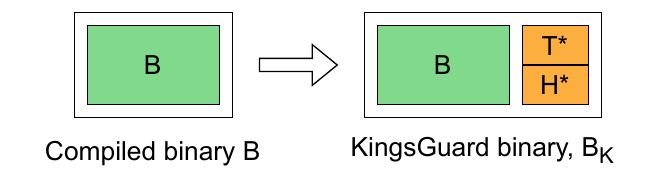}
    \caption{Taints $T^*$ and hashes $H^*$ added to the binary.}%
    \label{fig:binary_prep}
\end{figure}
\begin{figure}
    \centering
    \includegraphics[width=0.35\textwidth]{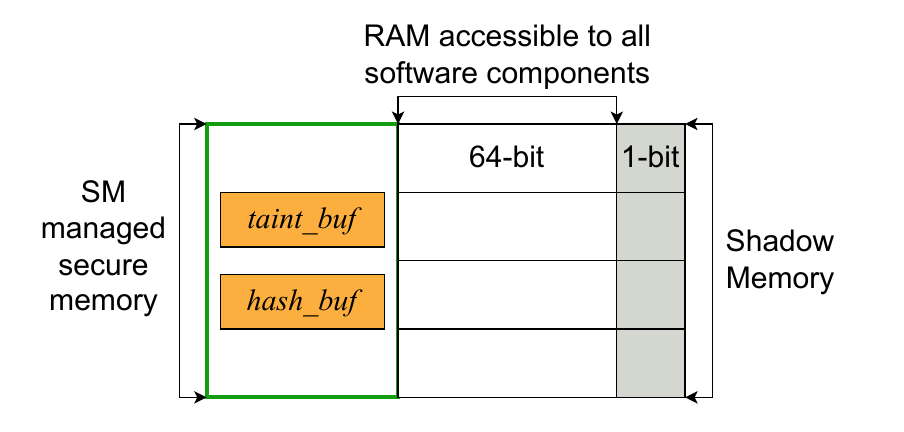}
    \caption{RAM structure in \kingsguard.}
    \label{fig:mem-struct}
\end{figure}
Unlike contemporary works~\cite{tarkhani2023deluminator, DBLP:journals/corr/abs-2401-08901, DBLP:conf/uss/TsaiSJMPP20}, \kingsguard does not need extensive modifications to the application source code. It works predominantly with the compiled enclave binary, embedding metadata necessary for fine-grained information flow tracking and controlled declassification. To achieve this, the enclave developer annotates sensitive data in the source code. 
This is done by assigning the sensitive data to a dedicated section in the binary by using the gcc section-name attribute (\texttt{\_\_attribute\_\_\allowbreak(section(".section\_name"))}). All data in this section is then associated with taints after compilation. For declassification, the developer identifies ADPs from the compiled binary by statically enumerating control flow paths that can legitimately release data from the enclave and computing hashes over them.

 The compiled binary, $B$, consists of multiple sections, including the dedicated section $D^*$ for sensitive data, which can be considered as a collection of 64-bit data words, $D$. A one-bit taint, $T$, is added to each 64-bit word that has been marked sensitive by the developer. All the taints are stored in a separate section of the binary, $T^*$, defined as:
\[
T^* = \{T~~ |~~ \forall~ D \in D^*\}.
\]


To support declassification, \kingsguard embeds hashes computed over the ADPs into the binary. For each identified ADP, $P \in P^*$, a cumulative hash $H$ is computed over the nodes $N^*$ in the execution path, where each node $N \in N^*$ represents a control-flow operation as a pair of source and target addresses ($s_i, t_i$). For loops, $N$ is represented as a pair of loop condition ($l_{c}$) and loop entry ($l_{e}$) addresses. Hash for a loop is computed once to ensures that deviations altering the loop structure (entry/exit) are detected. The ADPs embedded in the binary are a collection of such hashes, $H^*$, computed as follows:

\scalebox{0.85}{
\begin{minipage}{0.5\textwidth}
\begin{mathpar}
H_0 = hash(\mathrm{init} \,||\, s_0 \,||\, t_0),\\
\text{For } i = 1, 2, \ldots, n,~n = |N^*|: \\
H_i =
\begin{cases}
hash(H_{i-1} \,||\, s_i \,||\, t_i), & \text{branch} \\
hash(H_{i-1} \,||\, l_{c} \,||\, l_{e} ), & \text{loop}
\end{cases},\\
H^* = \{\, H \mid \forall P \in P^*: H = H_n \,\}
\end{mathpar}
\end{minipage}
}

The final binary, $B_K$, contains the existing sections in $B$ and the taint and hash sections ($T^*$, $H^*$), {\em i.e.} $B_{K} = (B, T^*, H^*)$ (see Figure~\ref{fig:binary_prep}). We assume the sections $T^*$ and $H^*$ are digitally signed and the signature is verified by the SM during load time to detect tampering.


{\flushleft \bf Loading the \kingsguard Binary.} When the OS encounters $T^*$ and $H^*$ while loading the enclave binary $B_K$, it transfers control to the SM, which verifies the integrity of these sections and stores them in protected buffers, $taint\_buf$ and $hash\_buf$, in a secure memory region accessible only to the SM (see Figure~\ref{fig:mem-struct}). The malicious OS may decide to not forward the taints and hashes to the SM, but the SM explicitly checks for their presence before enclave creation; if either section is missing, enclave creation is aborted.

\subsection{Baseline TEE Guarantees}~\label{sec:tee}

\begin{figure}[!t]
    \includegraphics[width=0.2\textwidth]{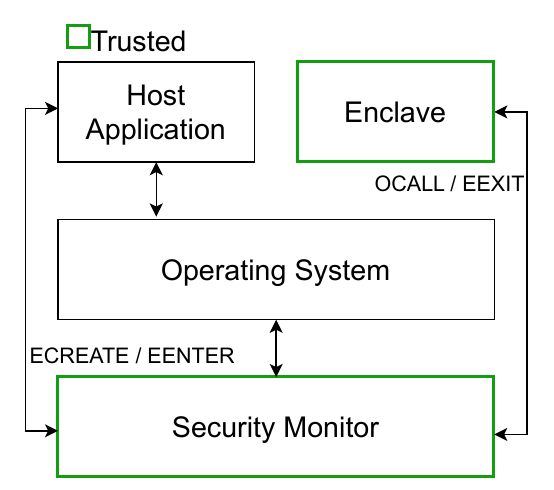}
    \caption{\kingsguard software stack. Enclave and SM are the only trusted software components. Host Application interacts with the enclave only via the SM.}
    \label{fig:tee}
\end{figure}
\begin{figure}
    \centering
    \includegraphics[width=0.25\textwidth]{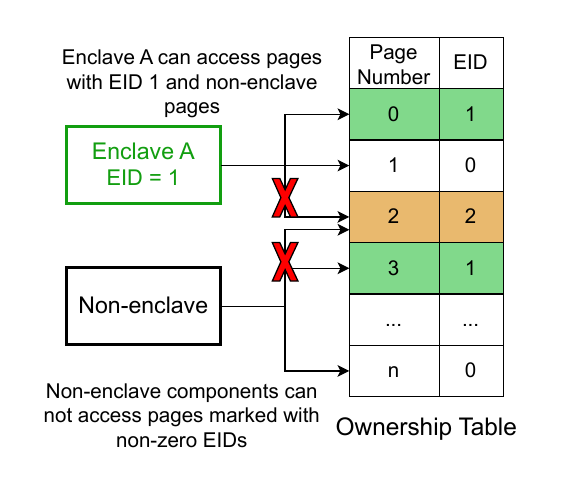}
    \caption{Non-enclave components are not allowed to access pages marked with an EID, while the enclave can access pages marked with its own EID and the unmarked pages.}
    \label{fig:ot}
\end{figure}
The taint tracking and declassification mechanism of \kingsguard can be applied to any generic TEE design including SGX~\cite{DBLP:journals/iacr/sgxCostanD16}, TrustZone~\cite{DBLP:conf/coinco/trustzoneNgabonzizaMBCM16}, and Confidential Virtual Machines (CVMs)~\cite{sev2020sevsnp, intel2022tdx, kaplan2016amd, armcca}. We adopt an SGX-style enclave design due to its well-established threat model and the extensive body of literature built around it. Further, SGX is the most targeted TEE for the attacks motivating this paper, making this design a natural and well-understood foundation for \kingsguard. 

\kingsguard isolates enclave code and data from all untrusted software components in the system. Non-enclave components can interact with the enclave via the SM that operates at a higher privilege level than the OS. The software stack for \kingsguard is shown in Figure~\ref{fig:tee}. The host application requests enclave creation by the SM ({\tt ECREATE}), which assigns a unique identifier, \emph{EID} to the enclave and records the enclave metadata. Physical memory is allocated to the enclave on demand when servicing page faults. Each physical page allocated to the enclave is dynamically added to the enclave memory by associating it with the enclave's EID. \kingsguard implements a dedicated hardware structure, called the \emph{Ownership Table} (OT), that maintains a mapping between physical pages and their corresponding EIDs (Figure~\ref{fig:ot}). Isolation is enforced in hardware by permitting memory access only when the EID of the executing enclave, $CurrEID$, matches the EID of the target page. The OS retains scheduling and paging control but cannot read or write to enclave memory.

Once the enclave is created and registered in the SM, the host application can request services from the enclave. This requires transitions into and out of the enclave, which are governed by the SM. The host application enters an enclave via the SM ({\tt EENTER}), which saves the host context and switches to the enclave context, transferring control to a fixed entry point. After completing its intended task, the enclave code exits via the SM ({\tt EEXIT}), which restores the host application context, transferring control back to the host application. Enclave code may also explicitly exit to request services like system calls. These transitions are also mediated by the SM ({\tt OCALL}), which copies arguments/results between enclave and non-enclave memory using SM-validated buffers or shared pages.

In addition to these intended transitions, enclaves may be asynchronously interrupted by hardware interrupts. These asynchronous exits ({\tt AEX}) are also redirected through the SM, which ensures a clean exit from the enclave by saving its context before transferring control to the interrupt handler. The execution inside the enclave can later be resumed from the saved context. To this baseline TEE, we incorporate information flow tracking and declassification in hardware, which permits secure operation despite vulnerabilities.

\subsection{Information Flow Tracking in Hardware}~\label{sec:dift}
\kingsguard prevents information leakage from enclaves using hardware-assisted dynamic information flow tracking. Sensitive data inside the enclave is marked with taints at compile time (\S~\ref{sec:bin_prep}). These taints are propagated with the data as it is operated upon. \kingsguard prevents any tainted data from leaving the enclave boundary without authorization. This section describes how these taints are stored in memory, propagated during execution, and verified at enclave exit.


{\noindent\textbf{Taint Storage.}}
\kingsguard implements a separate, dedicated memory region, called the \emph{shadow memory}, to store the taints corresponding to the contents of the main memory. Each 64-bit block in main memory is mapped to a 1-bit taint in shadow memory, as shown in Figure~\ref{fig:mem-struct}. The taints extracted from the enclave binary $B_K$ must be placed into the shadow memory locations that map to the physical addresses of the data once it is loaded in RAM. However, since the OS follows lazy loading, the physical pages are not allocated at program load time, but only on demand when a page fault occurs. So, the taints are placed in shadow memory when the corresponding data is loaded in RAM. To do this, \kingsguard intercepts page faults in the SM, checks if the faulting address belongs to the dedicated data section $D^*$, and then writes the taints into the shadow memory address corresponding to the physical address of the data loaded in RAM. 
Since tainting is done at a granularity of one bit per 64-bit word, each 4 KB page requires exactly 64 bytes of taints in shadow memory. This is illustrated in the following rule.

\scalebox{0.85}{
\begin{minipage}{0.5\textwidth}
\begin{mathpar}
\fbox{{\centering \textsc{ LOAD-PAGE $\langle~index~\rangle$}}} \\
\mbox{\ding{202}}~\dbox{\centering OS} \vartriangleright Pg_{num} = load\_page(index) \\
\mbox{\ding{203}}~\dbox{\centering SM} \vartriangleright Pg_{num} \in D^*\rightarrow \mbox{\ding{204}}~D^*_{off} = offset(Pg_{num})~in~D^*\\
\rule{0.8\textwidth}{0.4pt}
\\ \mbox{\ding{205}}~Shadow\_mem[index * 64] = taint\_buf[D^*_{off} \gg 6]
\end{mathpar} 
\end{minipage}
}

This rule explains the operations that occur during a page fault. \ding{202} When OS services a page fault, the return to the user space is intercepted in the SM. \ding{203} The SM checks if the faulting address is a data address and \ding{204} computes the offset of the faulting address in the data section of the enclave binary ($D^*_{off}$). \ding{205} The SM then updates the shadow memory address corresponding to the faulting address with the taint values present at the computed offset in $taint\_buf$.

{\noindent\textbf{Taint Propagation.}}\label{sec:taint-prop}
To ensure effective tracking of sensitive information, the taints associated with sensitive data must propagate as the data moves through the system. This includes tracking the taint when the data is loaded into processor registers, used in arithmetic or logical operations, or copied to other memory locations. To achieve this, it is essential to propagate the taints when data is loaded from memory into processor registers for computation. \kingsguard extends the register file by augmenting each register with a taint bit. For every data load operation, \kingsguard not only retrieves the data from memory, but also the associated taint from the shadow memory. This is illustrated in the following rule:  


\scalebox{0.85}{
\begin{minipage}{0.5\textwidth}
\begin{mathpar}
\fbox{{\centering \textsc{LOAD $\langle~r_v,r_a\rangle$}}} \\

\mbox{\ding{202}}~\dbox{\centering HW} \vartriangleright~\centering 
 ~val = mem(r_a) 
 \quad
  \mbox{\ding{203}}~taint = shadow\_mem(r_a) $\\$
\rule{0.8\textwidth}{0.4pt}
$\\$ \mbox{\ding{204}}~r_v = val \quad \mbox{\ding{205}}~r_t = taint
\end{mathpar}
\end{minipage}
}

This rule represents a load instruction in enclave code that loads a value $val$ from the address stored in register $r_a$ into a register $r_v$. \ding{202} The hardware reads $val$ from the address in $r_a$ and \ding{203} fetches the taint from the corresponding shadow memory region. \ding{204} It then updates $r_v$ to $val$ and \ding{205} the corresponding taint bit $r_t$ to $taint$.

\begin{figure}[!t]
    \centering \includegraphics[width=0.35\textwidth]{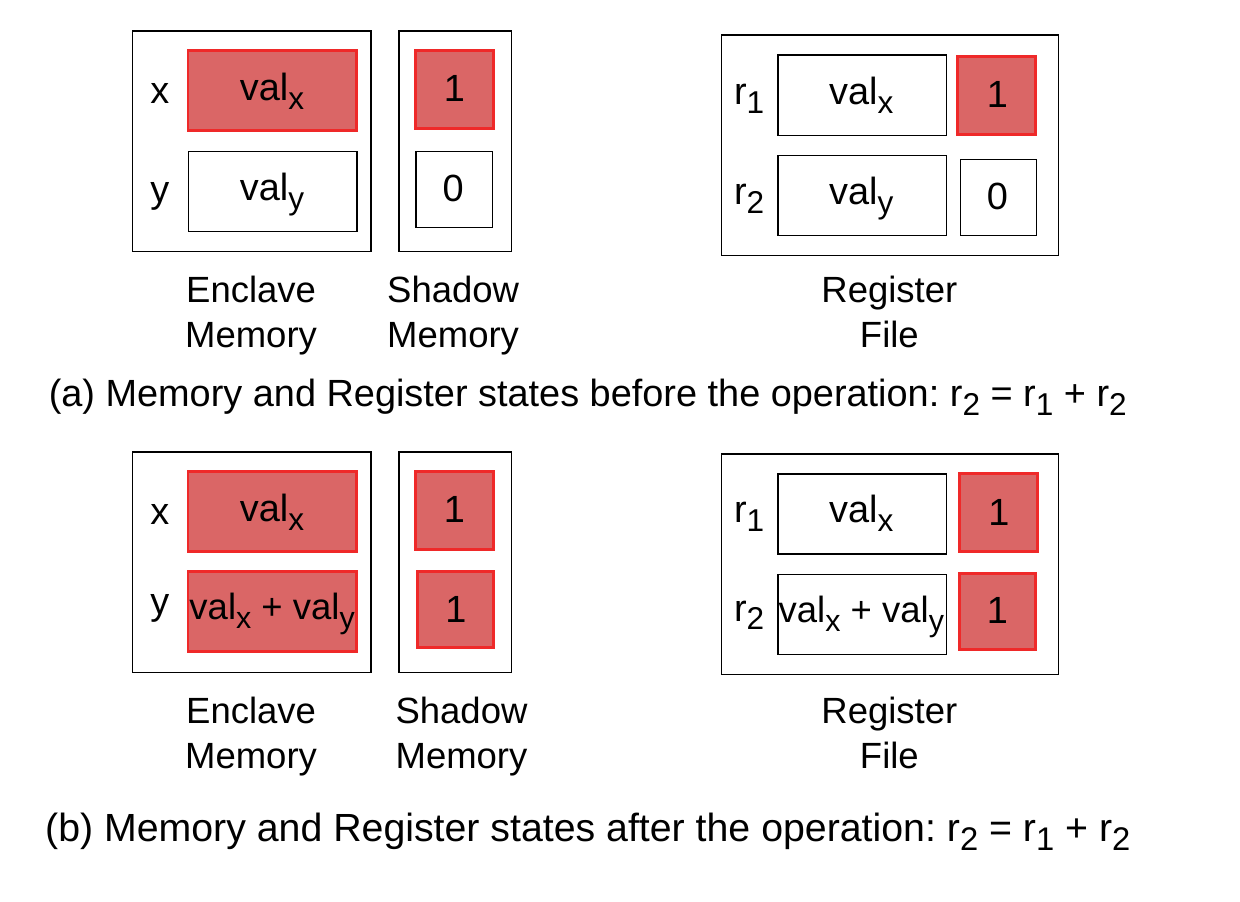}
    \caption{Output registers are tainted with the conservative union of source register taints.}
    \label{fig:taint-prop}
\end{figure}

Once present in the register, the taint must be propagated appropriately as the data is used in computations. This entails defining a comprehensive set of taint propagation rules that govern how taints are transferred, combined, or cleared depending on the semantics of the executed instruction. For example, in binary operations involving multiple operands, the output taint typically reflects a conservative union of the input taints, ensuring that any dependency on tainted inputs is preserved in the result, as demonstrated in Figure~\ref{fig:taint-prop}.
To formalize this behavior, \kingsguard specifies taint propagation semantics across different categories of instructions, such as arithmetic, logical, memory access, and control-flow operations. A summary of these rules is provided in Table~\ref{tab:taint_prop}. 

\begin{table*}[!t]
\centering
\caption{Taint Propagation Rules in \kingsguard}
\begin{adjustbox}{width=0.7\textwidth}
\begin{tabular}{llll}
\toprule
\textbf{Instruction Format} & \textbf{Fields} & \textbf{Operation} & \textbf{Taint Propagation}  \\
\midrule
Register operands & opcode, rd, rs1, rs2 & $rd \leftarrow rs1$ \textit{op} $rs2$ & $rd\_t \leftarrow rs1\_t \lor rs2\_t$ \\
Immediate operand & opcode, rd, rs1, imm & $rd \leftarrow rs1$ \textit{op} $imm$ & $rd\_t \leftarrow rs1\_t$ \\
Load & opcode, rd, rs1, imm & $rd \leftarrow mem(rs1 + imm)$  & $rd\_t \leftarrow shad\_mem(rs1 + imm)$ \\
Store & opcode, rs1, rs2, imm & $mem(rs2 + imm) \leftarrow rs1$  & $shad\_mem(rs2 + imm) \leftarrow rs1\_t$ \\
\bottomrule
\end{tabular}
\end{adjustbox}
\label{tab:taint_prop}
\end{table*}

While these propagation rules ensure accurate tracking of sensitive information within the enclave’s data path and memory hierarchy, information leakage can also occur through shared hardware resources. To prevent such leakage, \kingsguard extends the shared hardware registers, accessible in user space with an \emph{owner} field. The registers are tainted with the EID of the enclave that writes to it. This is illustrated in the following rule:

\scalebox{0.85}{
\begin{minipage}{0.5\textwidth}
\begin{mathpar}
\fbox{{\centering \textsc{WRITE-REG $\langle~r_s, val\rangle$}}} \\
\mbox{\ding{202}}~\dbox{\centering HW} ~~~ r_s \in U \quad \mbox{\ding{203}}~owner = CurrEID \\
\rule{0.8\textwidth}{0.4pt}
$\\$ \mbox{\ding{204}}~r_s = val \qquad \mbox{\ding{205}}~r_s.owner = owner
\end{mathpar}
\end{minipage}
}

This rule represents an operation that writes a value $val$ to a shared register $r_s$. \ding{202} The hardware checks if $r_s$ is a user space register. \ding{204} If so, it updates $r_s$ with $val$, and \ding{205} taints $r_s$ by updating its $owner$ field to the EID of the currently executing enclave, $CurrEID$.

{\noindent\textbf{Enforcing Taint Checks to Stymie Direct Copy.}}
While taint propagation enables effective tracking of sensitive data as it moves through the system, detecting actual data leakage requires checks at \emph{taint sinks}, the points where sensitive data could potentially escape the enclave. In \kingsguard, non-enclave memory is treated as the primary taint sink. Thus, every non-enclave memory access from the enclave is subjected to a hardware-level check to determine if tainted data is being written out. On a store operation, the taint associated with the data being written out is inspected. If the data is tainted, \kingsguard prevents it from being written to non-enclave memory unless authorized through declassification (\S~\ref{sec:declass}. This is illustrated in the rule \textsc{STORE-MEM}, where a store operation is trying to write the value in register $r_v$ to an address in $r_a$. \ding{202} The hardware checks if the address in $r_a$ belongs to non-enclave memory. \ding{203}, \ding{204} If so, it checks the taint associated with the register $r_v$ and if the register is tainted ($r_t == 1$), \ding{205} sensitive data leakage is prevented by replacing the contents of the tainted register with zeroes.

\scalebox{0.85}{
\begin{minipage}{0.5\textwidth}
\begin{mathpar}
\fbox{{\centering \textsc{STORE-MEM $\langle~r_a, r_v\rangle$}}} \\
\mbox{\ding{202}}~\dbox{\centering HW} ~~~ r_a \in non\_enclave\_mem \quad \mbox{\ding{203}}~r_t = taint(r_v) \quad \mbox{\ding{204}}~r_t == 1 \\
\rule{0.8\textwidth}{0.4pt}
$\\$ \mbox{\ding{205}}r_v = 0 \quad r_t = 0 \quad mem(r_a) = 0
\end{mathpar}
\end{minipage}
}

{\noindent\textbf{Enforcing Taint Checks to Stymie Indirect Leakage.}}
\kingsguard also enforces checks on accesses to tainted addresses in non-enclave memory to prevent attackers from encoding secrets as memory addresses. This is illustrated in the rule \textsc{ACCESS-MEM}, which represents a memory operation accessing an address stored in register $r_a$. \ding{202} If the address register points to non-enclave memory, \ding{203}, \ding{204} and is tainted, \ding{205} access to this address is prevented by replacing it with a fixed valid address $a_{fixed}$ in non-enclave memory.

\scalebox{0.85}{
\begin{minipage}{0.5\textwidth}
\begin{mathpar}
\fbox{{\centering \textsc{ACCESS-MEM $\langle~r_a\rangle$}}} \\
\mbox{\ding{202}}~\dbox{\centering HW} ~~~ r_a \in non\_enclave\_mem \quad \mbox{\ding{203}}~r_t = taint(r_a) \quad \mbox{\ding{204}}~r_t == 1 \\
\rule{0.8\textwidth}{0.4pt}
$\\$ \mbox{\ding{205}}~r_a = a_{fixed}
\end{mathpar}
\end{minipage}
}


To prevent data leakage via shared hardware registers, \kingsguard enforces checks as illustrated in the following rule:

\scalebox{0.85}{
\begin{minipage}{0.5\textwidth}
\begin{mathpar}
\fbox{{\centering \textsc{READ-REG $\langle~r_s\rangle$}}} \\
\mbox{\ding{202}}~\dbox{\centering HW} ~~~ r_s \in U \quad \mbox{\ding{203}}~owner = r_s.owner \quad \mbox{\ding{204}}~CurrEID == owner \\
\rule{0.8\textwidth}{0.4pt}
$\\$ \mbox{\ding{205}}~val = r_s
\end{mathpar}
\end{minipage}
}

This rule represents a read operation from a shared register $r_s$. \ding{203} The hardware checks if $r_s$ is a user-space register and \ding{203}, \ding{204} compares the $owner$ field of $r_s$ with $CurrEID$. \ding{205} The read access is allowed only when the two match.
This mechanism ensures that shared hardware state cannot be used to exfiltrate data from enclaves.
\subsection{ Declassification for Authorized Data Flows }~\label{sec:declass}
\kingsguard allows legitimate data release only through an ADP. 
To ensure data is released from the enclave only through ADPs, \kingsguard computes a cumulative hash~$H_{current}$ for branch instructions at runtime similar to $H_i$ in Section~\ref{sec:bin_prep}. 
Whenever a store instruction to non-enclave memory is encountered, hardware verifies if $H_{current} \in H^*$; only if this constraints hold, tainted data is declassified. This is illustrated as follows:

\scalebox{0.85}{
\begin{minipage}{0.5\textwidth}
\begin{mathpar}
\fbox{{\centering \textsc{DECLASSIFY $\langle~r_a, r_v\rangle$}}} \\
\mbox{\ding{202}}~\dbox{\centering HW} ~~~ r_a \in non\_enclave\_mem 
\quad 
\mbox{\ding{203}}~r_t = taint(r_v) 
\quad 
\mbox{\ding{204}}~r_t == 1 \\
\mbox{\ding{205}}~\exists H \in H^*~:~H_{current} == H \\
\rule{0.8\textwidth}{0.4pt} \\
\mbox{\ding{206}}~r_t = 0 
\quad 
\mbox{\ding{207}}~mem(r_a) = r_v
\end{mathpar}
\end{minipage}
}

In this rule, \ding{202}, \ding{203}, \ding{204}, hardware detects a store operation writing tainted data in $r_v$ to a non-enclave memory address $r_a$, and then \ding{205} checks if $H_{current}$ matches any hash present in $H^*$. \ding{206} If the hash matches, the taint associated with the data is removed, and \ding{207} the store operation succeeds.


%% file: Implementation.tex
We have implemented \kingsguard on the Shakti-C class processor~\cite{veezhinathan2022building}, which is a 5-stage in-order RISC-V processor and synthesized the design on a Xilinx Arty A7-100T FPGA board. The CPU features separate 16 KB, 4-way set-associative L1 instruction and data caches with a 64-byte cache line size. The L1 caches are connected via an AXI-4 system bus to a Xilinx memory controller, which interfaces with DDR3 memory on the FPGA board. This section discusses the implementation aspects of \kingsguard.




\subsection{Implementing TEE Primitives}

\begin{figure}[!t]
    \includegraphics[width=0.45\textwidth]{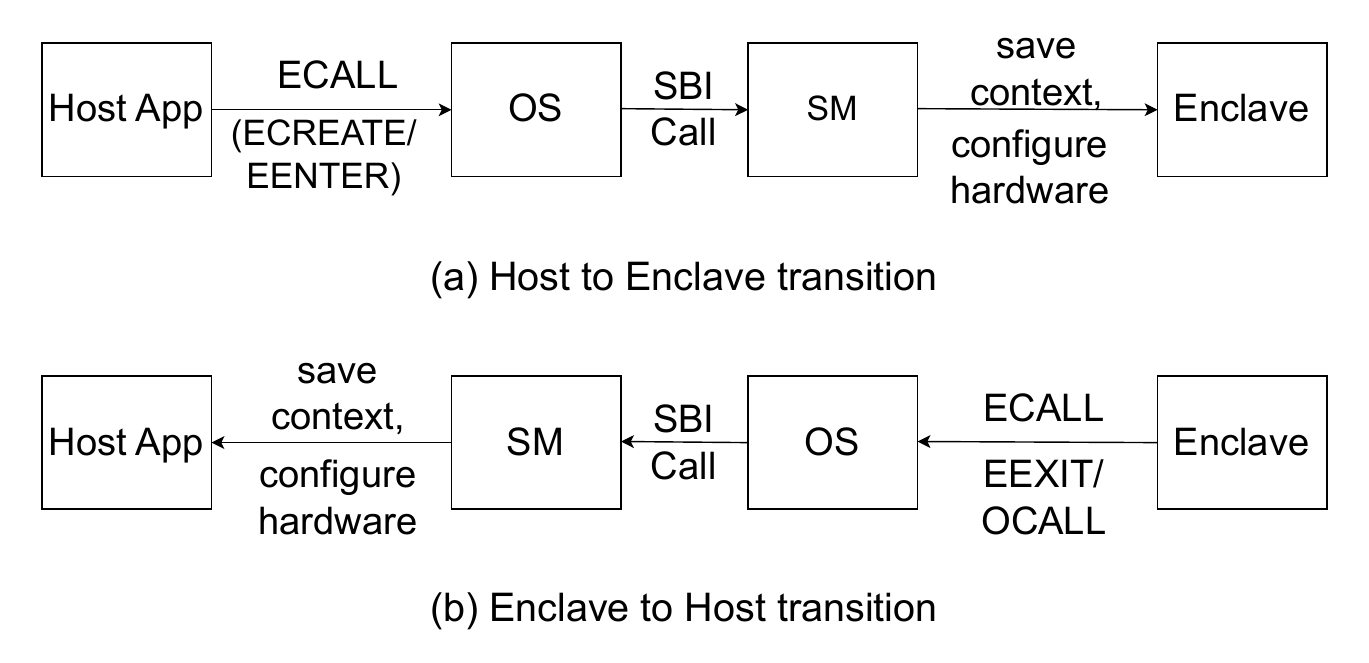}
     \caption{Interaction between enclave and non-enclave components using ECALLs.}
    \label{fig:ecall}
\end{figure}
\begin{figure}
    \centering
    \includegraphics[width=0.3\textwidth]{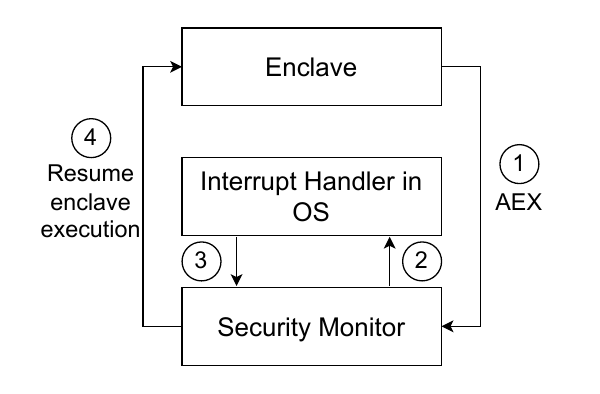}
     \caption{Handling of asynchronous interrupt in an enclave.}
    \label{fig:aex}
\end{figure}

To create and execute an enclave, \kingsguard introduces four enclave management calls, called ECALLs: {\tt ECREATE}, {\tt EENTER}, {\tt EEXIT}, and {\tt OCALL}. These calls reuse Intel's SGX nomenclature and serve as the software interface between the host application and the enclave. In our implementation on the RISC-V architecture, these ECALLs are realized using the standard \texttt{ecall} instruction, which triggers an \emph{environment call exception}. When this instruction is executed in User mode, control is transferred to the Supervisor mode, allowing the operating system to handle the request.
 In \kingsguard, when a user invokes any of these ECALLs, the CPU core raises an \texttt{ecall} trap in Supervisor mode, where the OS identifies these ECALLs and redirects them to the SM via an SBI (Supervisor Binary Interface) call. The SM has a trap handler for each one of these calls. The transitions using these calls are shown in Figure~\ref{fig:ecall}.

{\noindent\textbf{ECREATE:}} When the host application invokes {\tt ECREATE}, the SM creates the enclave by assigning it a unique EID and registering its entry point address.

{\noindent\textbf{EENTER:}} To execute enclave code, the host application calls {\tt EENTER}. On {\tt EENTER}, the SM saves the host context, sets the value of the Program Counter (PC) to the enclave's entry point address and configures hardware registers to indicate that the core is now executing in enclave mode. This includes enabling taint tracking and hashing in hardware. It also updates the $CurrEID$ register in hardware to the EID of the enclave.

{\noindent\textbf{EEXIT:}} Once the enclave finishes its task, it invokes the {\tt EEXIT} call. The SM clears the general-purpose registers used by the enclave, resets hardware configuration registers to enable execution in non-enclave mode, updates $CurrEID$ to 0, restores the host context, and updates the PC to host application instruction after EENTER. Taint tracking and hashing are disabled on exit from enclave mode.

{\noindent\textbf{OCALL:}} To request services from the OS, enclave code calls {\tt OCALL} with the necessary data that needs to be copied outside the enclave. SM copies this data from the enclave to non-enclave memory after validating the size and then transfers control to the host application in a similar way to {\tt EEXIT}. However, the SM records the PC making the {\tt OCALL} to later resume execution at this address. The SM also copies the results from the host back into the enclave memory after proper bounds checking.

{\noindent\textbf{AEX:}} During enclave execution, interrupts are redirected to the SM instead of the OS. \emph{AEX\_handler} is implemented in the SM, which saves the enclave context in a secure memory region, clears the registers, configures hardware registers to run the core in non-enclave mode, and then transfers control to the actual interrupt handler. After the execution of the interrupt handler, control is transferred back to the SM, which restores the enclave context and resumes execution (Figure~\ref{fig:aex}).

{\noindent\textbf{Enclave Isolation:}} Enclave isolation in \kingsguard is enforced by the \emph{Ownership Table} (OT) implemented in hardware. The OT is a vector of 64-bit registers, where each entry maps to a physical page in RAM. The entries in the OT are initialized to zero, indicating that the page can be accessed by any process. When a physical page is allocated to an enclave, its OT entry is updated to the EID of the enclave. The pages marked with an EID can only be accessed by the enclave with the same EID.

\subsection{Implementing Information Flow Tracking}
To support fine-grained tracking of sensitive data in hardware, \kingsguard implements a \emph{shadow memory} region to store taint bits corresponding to every data block in memory and extends the register file to associate each register with a taint bit. These components enable continuous tracking as sensitive data propagates through memory and registers during enclave execution.

{\noindent\textbf{Shadow Memory:}} The shadow memory is a reserved region in physical memory used to store the taint bits for all data blocks in main memory. This region of memory is not accessible to the OS or user processes. This ensures that the taints remain secure and untampered. In our implementation, each 64-bit data block in the main memory is associated with a single 1-bit taint. The mapping between data and its corresponding taint bit is computed using a simple linear address translation:
\[
\boxed{
\begin{aligned}
\texttt{taint\_address} &= \texttt{((data\_address - data\_memory\_base)}\\
&\quad \texttt{>> 6) + shadow\_memory\_base}
\end{aligned}
}
\]

 On every memory load request, hardware computes the \texttt{taint\_\allowbreak address} and sends an additional memory request to fetch the taint along with the data. Similarly, on a store request, the taint is written back to the shadow memory. These additional taint requests are generated only when the processor is executing in enclave mode. In our implementation, the data and taints share the same cache hierarchy to simplify coherence management. This design choice eliminates the need for a separate cache for taints, reducing hardware complexity while maintaining performance efficiency.

{\noindent\textbf{Register File:}} The register file is extended to store 1-bit taint for every 64-bit register. The interface to the register file is also extended with methods to read and write the taints. Pipeline stages use these interfaces to fetch and update a register’s taint alongside its value, enabling fine-grained information-flow tracking at the register level.

{\noindent\textbf{Taint Propagation:}} To propagate taints, the instruction decoder and execution pipeline are extended to handle taint metadata in parallel with normal instruction execution. For every arithmetic, logical, load, or store operation, \kingsguard updates the output taint based on the propagation rules defined in Table \ref{tab:taint_prop}. 

\subsection{Implementing Declassification}

To authorize data leaving the enclave, \kingsguard maintains a cumulative runtime hash,~$H_{current}$, in the hardware. It is computed by a dedicated hardware component in the writeback stage of the processor pipeline. The component is enabled at the start of enclave execution and comprises of a Branch Monitor and a Hash Engine. On every instruction commit, the control flow instructions (branches and jumps) are identified by the Branch Monitor, which determines the source address {$s_i$} (See \S~\ref{sec:bin_prep}) from the current PC. The target address {$\tt t_i$} is derived from the PC of the subsequent instruction commit after a branch or jump instruction. {$s_i$ and $t_i$} are then sent to the Hash Engine, which computes the running hash $H_{current}$ as discussed in Section~\ref{sec:declass}.
Loops in the program execution are detected by a conditional branch with a negative offset. When a loop is found, the hardware records the addresses corresponding to the loop condition instruction and the loop entry as discussed in Section~\ref{sec:bin_prep} to compute ~$H_{current}$.  

Hash computation is performed by a SHA-2 engine that generates a 256-bit hash value stored in the $H_{current}$ register. 
On {\tt EEXIT}, the Hash Engine is disabled and the $H_{current}$ register is reset. However, on {\tt AEX}, only the Hash Engine is disabled, while maintaining the value in $H_{current}$ register, so that it can continue to be used for the hash computation once enclave execution resumes.

%% file: Results.tex

We evaluate \kingsguard using a RISC-V processor synthesized on a Xilinx ARTY A7-100T FPGA. We also extend the cycle-accurate in-order Minor CPU model in the Gem5 simulator~\cite{DBLP:journals/sigarch/BinkertBBRSBHHKSSSSVHW11} with \kingsguard. The system runs Linux kernel version 5.10 to emulate a realistic runtime environment. The untrusted operating system runs in supervisor mode (S-mode), SM operates in machine mode (M-mode) using the RISC-V Proxy Kernel and the enclave application executes in user mode (U-mode). 
This section presents a detailed evaluation of \kingsguard in terms of both design and performance overheads. We demonstrate its practical effectiveness through a case study based on a SCADA Application in Section~\ref{sec:scada}.

\subsection{Design Overheads}
We quantify the design overheads of \kingsguard as hardware and software changes.
{\flushleft\textbf{Hardware.}} Table~\ref{tab:hw_overhead} shows the hardware  overhead incurred by implementing \kingsguard\footnote{The results are obtained from Xilinx's Vivado tool version v2019.2.}. The overhead in the TEE implementation stems primarily from the implementation of Ownership Table. For Information Flow Tracking (IFT), overheads arise due to extending the register file and tracking the taints. The declassification mechanism contributes most to the hardware area overhead due to the inclusion of a dedicated branch monitor, along with a hardware hashing engine. Together, these modules require 2,233 LUTs and 4,098 registers, of which the hashing engine alone accounts for approximately 93\% of the LUTs and 95\% of the registers. The significant contribution of the hashing unit stems from the integration of a SHA-2 hardware core to ensure efficient and secure runtime verification. Importantly, these changes do not impact the processor’s maximum operating frequency but increase the power consumption by 0.9\%. These results are comparable to existing RISC-V based TEEs~\cite{DBLP:conf/uss/cureBahmaniBDJKSS21, allaqband2025tesla}.

{\flushleft\textbf{Software.}} Software changes for \kingsguard are quantified in Table~\ref{tab:sw_overhead}. Other than annotating sensitive data and ADPs, \kingsguard does not require any changes in the application. It requires 90 Lines of Code (LOC) added to the Linux kernel to identify each ECALL and transfer control to the SM. The OS also needs to load the enclave binary on ECREATE, identify the taint and hash sections in the binary and pass these to the SM. The SM has five SBI calls, one for each ECALL, and one to load the taints and hashes. These, along with the {\tt AEX} and page fault handler, account for 341 LOC.





\noindent
\begin{minipage}[t]{0.65\columnwidth}
\scriptsize
\captionof{table}{Hardware overhead of \kingsguard for each component compared to the unmodified baseline processor.}
\begin{tabular}{@{}lccccc@{}}
\toprule
\textbf{Configuration} & \textbf{LUTs} & \textbf{Registers} \\
\midrule
Baseline &
47297 & 43441\\
TEE &
+985 & +837\\
IFT &
+1812 & +293 \\
Declassification &
+2356 & +4491\\
\kingsguard &
52450 (10.8\%) & 49062 (12.9\%)\\
\bottomrule
\end{tabular}

\label{tab:hw_overhead}
\end{minipage}
\hspace{1mm}
\begin{minipage}[t]{0.3\columnwidth}
\scriptsize
\captionof{table}{LOC added for different components in \kingsguard.}
\begin{tabular}{@{}ll@{}}
\toprule
\textbf{Component} & \textbf{LOC} \\
\midrule
Linux Kernel & 90 \\
Security Monitor & 341 \\
\bottomrule
\end{tabular}

\label{tab:sw_overhead}
\end{minipage}

\subsection{Performance Overheads}

\begin{figure}[!t]
    \includegraphics[width=0.35\textwidth]{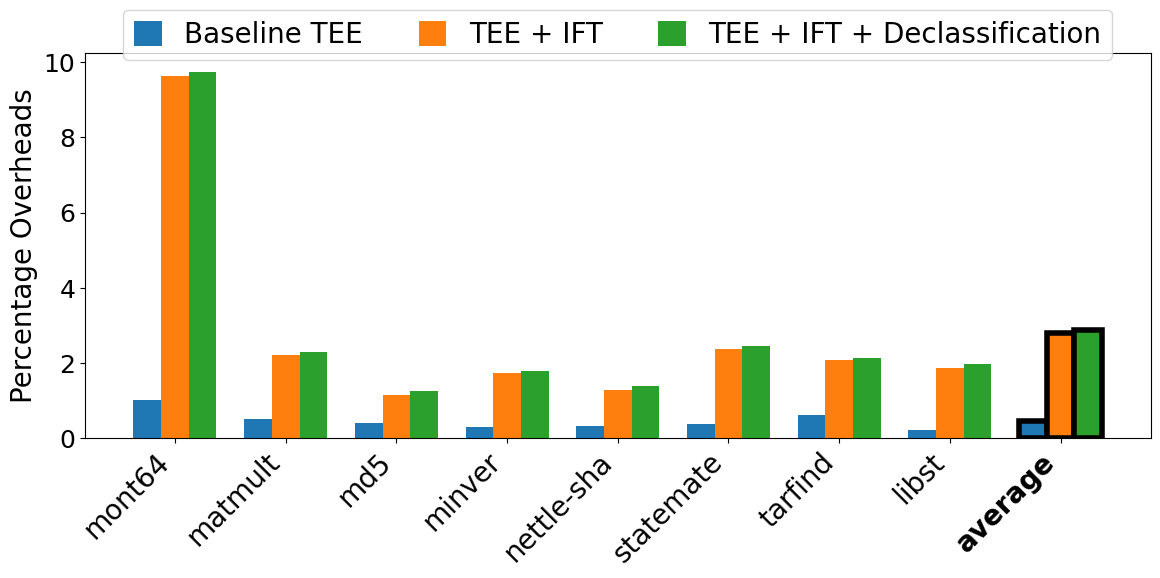}
     \caption{Percentage overheads of \kingsguard implemented as a 1) Baseline TEE, 2) TEE with Information Flow Tracking, 3) TEE with Information Flow Tracking and Declassification evaluated using Embench~\cite{patterson2023embench}.}
    \label{fig:embench}
\end{figure}

\begin{figure}[!t] 
  \centering
  \begin{minipage}{0.5\textwidth}
    \centering
    {\kingsguard's:\hspace{1.0em}Baseline TEE}\hspace{1mm}\legenditem{basicTEE}\hspace{1.8em}
    {TEE + IFT}\hspace{1mm}\legenditem{teeIFT}\hspace{1.8em}
    {TEE + IFT + Declassification}\hspace{1mm}\legenditem{teeIFTDec}
  \end{minipage}
  \begin{subfigure}[t]{0.49\linewidth}
    \centering
    \includegraphics[width=\linewidth]{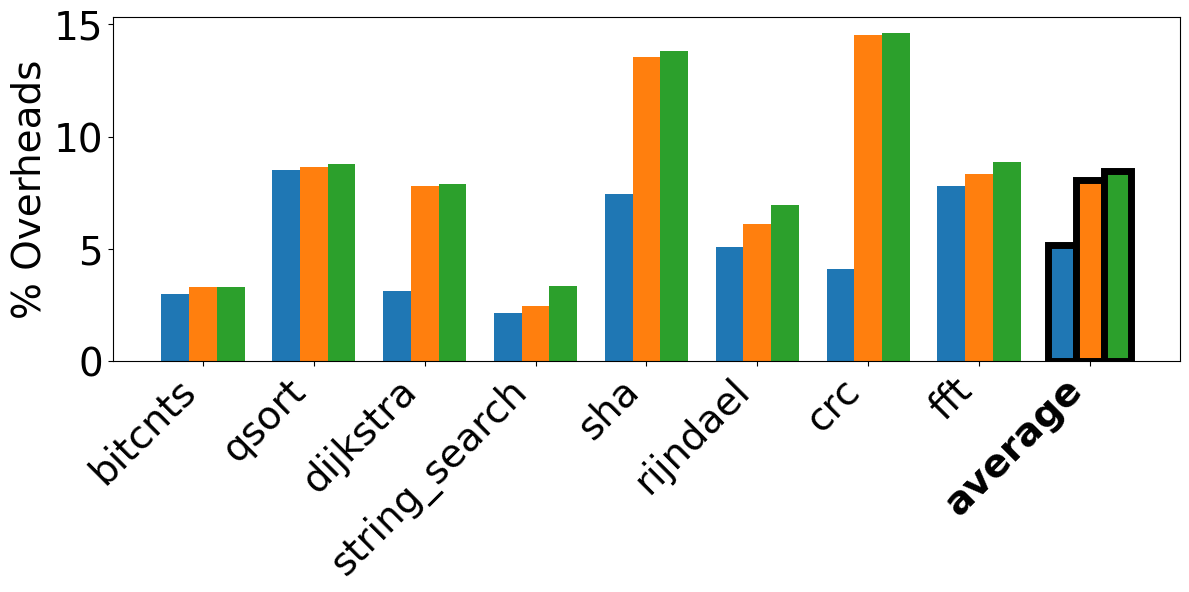}
    \caption{L1 Data Cache: 16 KB, L2 Cache: 128 KB}
    \label{fig:a}
  \end{subfigure}\hfill
  \begin{subfigure}[t]{0.49\linewidth}
    \centering
    \includegraphics[width=\linewidth]{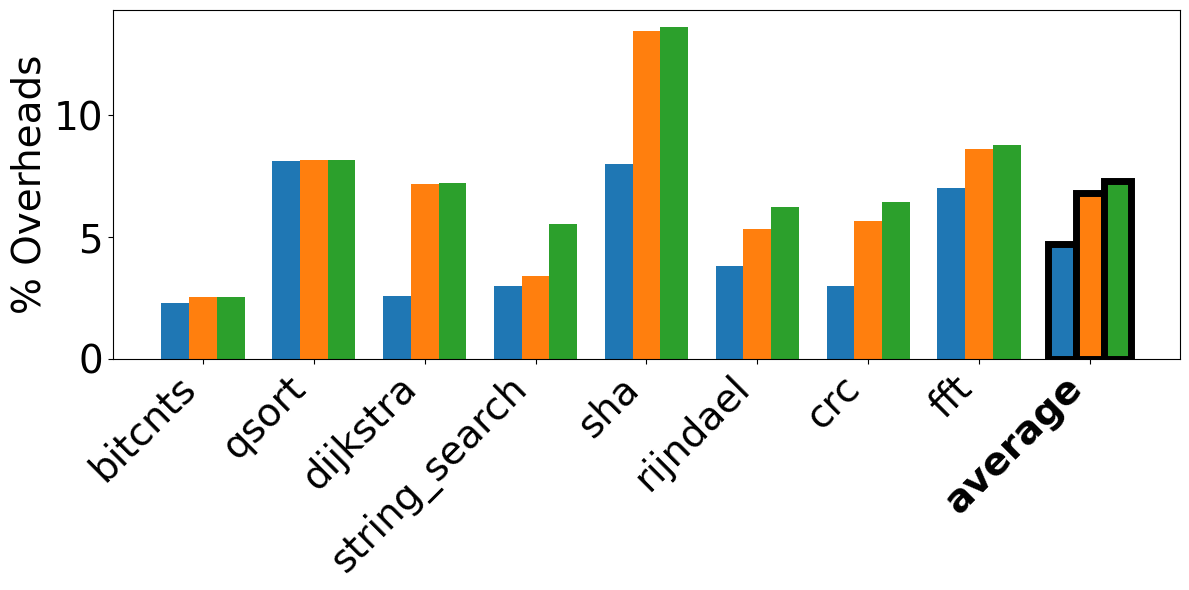}
    \caption{L1 Data Cache: 32 KB, L2 Cache: 256 KB}
    \label{fig:b}
  \end{subfigure}

  \vspace{0.75\baselineskip}

  \begin{subfigure}[t]{0.49\linewidth}
    \centering
    \includegraphics[width=\linewidth]{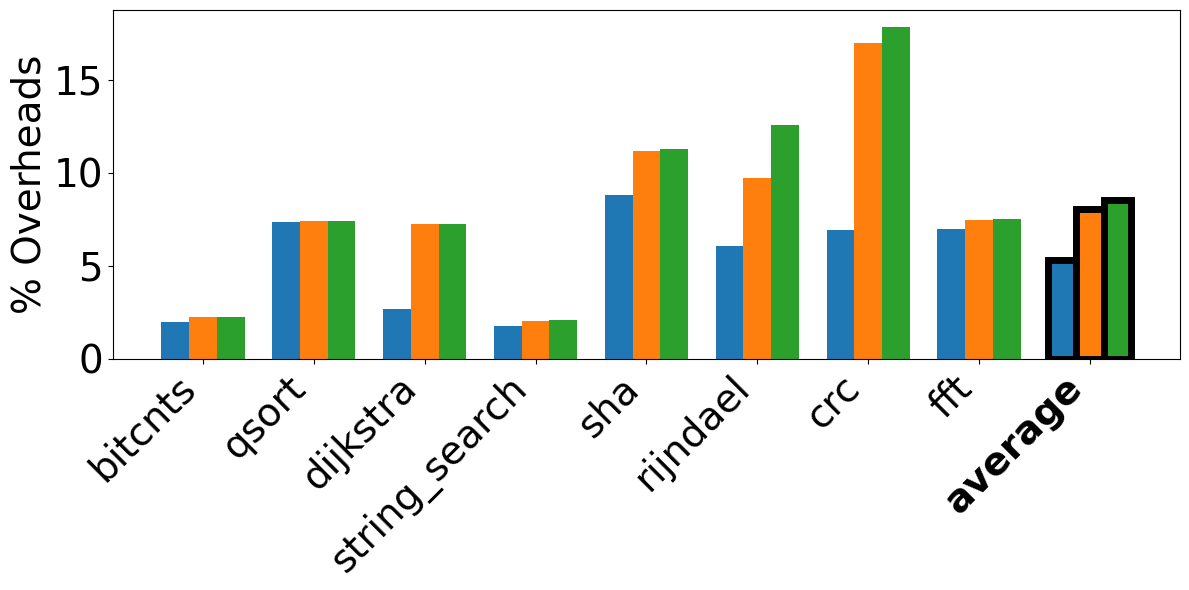}
    \caption{L1 Data Cache: 64 KB, L2 Cache: 256 KB}
    \label{fig:c}
  \end{subfigure}\hfill
  \begin{subfigure}[t]{0.49\linewidth}
    \centering
    \includegraphics[width=\linewidth]{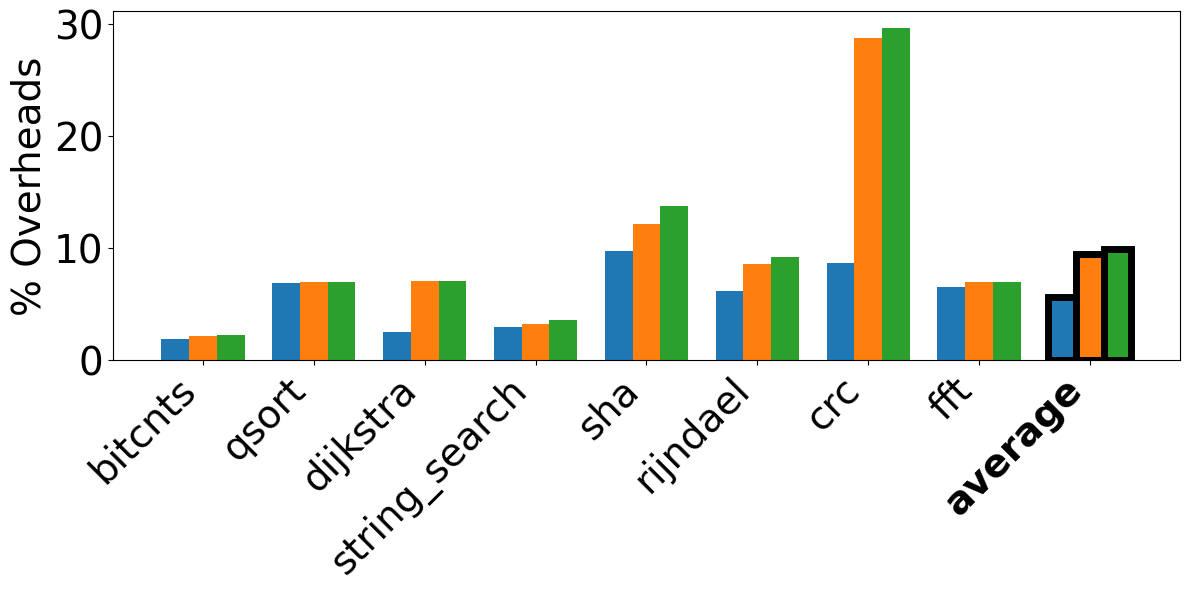}
    \caption{L1 Data Cache: 64 KB, L2 Cache: 512 KB}
    \label{fig:d}
  \end{subfigure}
 
  \caption{Performance overheads of \kingsguard for different cache sizes using MiBench~\cite{guthaus2001mibench}.}
  \label{fig:perf-over}
\end{figure}

We evaluate the performance overheads of \kingsguard using Embench~\cite{patterson2023embench} and MiBench~\cite{guthaus2001mibench} benchmark suites.
Embench is a modern, open-source benchmark suite that stress tests diverse aspects of the microarchitecture, including integer arithmetic, memory access patterns, and control flow, while MiBench is a widely used embedded benchmark suite~\cite{allaqband2025tesla, DBLP:conf/sp/GinerSPEUMG23} that models realistic application workloads.
Figure~\ref{fig:embench} shows the overheads of \kingsguard using Embench with respect to a non-enclave process. We evaluate overheads in three cases: 1) \kingsguard with only the baseline TEE features, 2) \kingsguard with TEE and IFT, and 3) \kingsguard with TEE, IFT, and declassification. The performance is evaluated with the entire program running inside the enclave. The baseline TEE adds around 0.46\% overhead on average. IFT introduces an additional 2.33\% overhead, while declassification adds a minimal overhead of 0.09\%. Except for \texttt{mont64}, most Embench benchmarks exhibit low IFT overhead. This is because \texttt{mont64} performs significantly higher percentage of memory accesses than the other workloads. While IFT increases the total number of memory accesses for all benchmarks due to shadow-memory operations, the resulting cache contention is substantially higher for memory-intensive programs like \texttt{mont64}, leading to an increased cache miss rate and higher performance overhead. The overhead due to declassification depends on the number of control flow instructions in the programs since hashing triggers selectively on control-flow instructions, and not for every instruction. The baseline TEE overheads depend on the number of context switches between enclave and non-enclave code.

Figure~\ref{fig:perf-over} shows the performance overheads of \kingsguard using {\tt MiBench} evaluated for different cache sizes to analyze how sharing of cache between the data and taints affects performance. The baseline TEE adds around 5\% overhead on average. IFT introduces an additional 2.8\% overhead, while declassification adds a minimal overhead of 0.5\%. Since each memory access triggers a parallel taint access and both data and taints share the same caches, smaller caches experience increased contention and higher miss rates, leading to degraded performance. Conversely, larger caches can accommodate both data and taints more effectively, thereby improving performance. This is clear from the overheads of {\tt sha} in Figure~\ref{fig:perf-over}. As the cache sizes increase, the overhead reduces from 6.1\% to 2.3\%. For most benchmarks, the performance impact remains minimal and largely consistent across cache configurations. However, memory-intensive workloads such as {\tt dijkstra} exhibit higher overhead due to frequent memory accesses and limited cache reuse. Interestingly, {\tt crc} demonstrates non-monotonic behavior. The overhead initially decreases with increasing the cache sizes, but rises again as the cache capacity increases further. This can be attributed to the initial benefit of improved taint–data cache locality, followed by diminished returns as the baseline program itself begins to exploit cache locality more effectively, which gets affected due to the introduction of taints in the cache.


\subsection{SCADA-based Partial Discharge Monitoring in Transformers }\label{sec:scada}
SCADA applications are security-critical and are prone to attacks~\cite{doj_colonial_pipeline_2021}. For this reason, these applications increasingly use TEEs to provide protection. However, if the application itself contains a vulnerability, such as a buffer overflow, it can still leak data despite being implemented within a TEE. To demonstrate this, we use an example SCADA application for detecting Partial Discharge (PD) in transformers~\cite{8984255} and introduce a buffer overflow vulnerability that leaks unencrypted data from the enclave. \kingsguard is able to prevent this leakage.

Time-series PD data are collected from local sensors on transformers, encrypted at the source of acquisition, and transmitted to a central monitoring system. Within this system, a Frequency Domain Analysis (FDA) module is deployed inside an enclave protected by \kingsguard. The encrypted time-series data is decrypted only within the enclave, where the FDA computation is securely performed. Transformed data produced by the FDA is re-encrypted before it leaves the enclave. The buffers that store the encrypted data are tainted a priori during the binary preparation stage, and the ADP is computed along the encryption path. 
\kingsguard adds an overhead of 5.55\%, of which baseline TEE contributes 3.1\%, IFT 2.25\%, and declassification 0.2\%.

 To evaluate \kingsguard’s effectiveness, we introduce a memory-corruption vulnerability in the FDA enclave module that can be exploited to leak unencrypted data from the enclave. We run the same on gem5.
 When the attacker attempts to write tainted unencrypted data to non-enclave memory, $H_{current}$ is checked against the ADP. Due to a mismatch, the store operation is prevented, thereby preventing the data leak. We also modify the FDA module to write the decrypted data to a user-accessible register that we have added to the processor to demonstrate that \kingsguard can prevent indirect data leaks via shared hardware registers. In an unprotected scenario, once the context switches to non-enclave code, it is able to read this value from the register. However, \kingsguard stamps the register with the EID of the enclave and prevents the non-enclave code from reading it even after the context switch.

%% file: Security.tex
\newcommand{\circlesize}{0.12cm}


\DeclareRobustCommand{\pre}{\tikz[baseline=-0.6ex]{\fill[black] (0,0) circle (\circlesize);}} 

\DeclareRobustCommand{\detect}{\tikz[baseline=-0.6ex]{ \filldraw[amber] (0,\circlesize) arc(90:-90:\circlesize) -- cycle; \draw[black] (0,0) circle (\circlesize); }}

\DeclareRobustCommand{\parpre}{\tikz[baseline=-0.6ex]{ \filldraw[black] (0,-\circlesize) arc(-90:90:\circlesize) -- cycle; \draw[black] (0,0) circle (\circlesize); }} 
\DeclareRobustCommand{\nodetect}{\tikz[baseline=-0.6ex]{\draw[black, thick] (0,0) circle (\circlesize);}}
\begin{table}[!t]
\centering
\caption{
Data Leak Attacks vs.~Defenses
($\pre$:~Prevents the leak, 
$\parpre$:~Partially prevents leak,
$\detect$:~Detects the leak, 
$\nodetect$:~Cannot Protect/Detect)}
\resizebox{\columnwidth}{!}{%
\begin{tabular}{|l|l|c|c|c|c|c|c|c|}
\hline
\multicolumn{1}{|c|}{\multirow{3}{*}{\textbf{AV}}} &
\multicolumn{1}{c|}{\multirow{3}{*}{\textbf{Data Leak Attacks}}} &
\multicolumn{7}{c|}{\textbf{Defenses}} \\ \cline{3-9}

\multicolumn{1}{|c|}{} & \multicolumn{1}{c|}{} &
\multicolumn{4}{c|}{\textbf{Software}} &
\multicolumn{3}{c|}{\textbf{Hardware}} \\ \cline{3-9}

\multicolumn{1}{|c|}{} & \multicolumn{1}{c|}{} &
\begin{tabular}[c]{@{}c@{}}GuaranTEE\\ \cite{morbitzer2023guarantee}\end{tabular} &
\begin{tabular}[c]{@{}c@{}}SGXMonitor\\ \cite{DBLP:conf/acsac/ToffaliniP0C22}\end{tabular} &
\begin{tabular}[c]{@{}c@{}}Deluminator\\ \cite{tarkhani2023deluminator}\end{tabular} &
\begin{tabular}[c]{@{}c@{}}HasTEE+\\ \cite{DBLP:journals/corr/abs-2401-08901}\end{tabular} &
\begin{tabular}[c]{@{}c@{}}Rezone\\ \cite{cerdeira2022rezone}\end{tabular} &
\begin{tabular}[c]{@{}c@{}}Light-Enclave\\ \cite{Gu2022AHC}\end{tabular} &
\kingsguard \\ \hline

\multirow{9}{*}{AV1}
 & CVE-2015-6639~\cite{beniamini2015android}
 & \detect & \detect & \detect & \pre
 & \parpre & \parpre & \pre \\ \cline{2-9}

 & CVE-2016-2431~\cite{beniamini2015trustzone}
 & \detect & \detect & \detect & \pre
 & \parpre & \parpre & \pre \\ \cline{2-9}

 & Pointer-Based~\cite{stella}
 & \detect & \detect & \detect & \pre
 & \nodetect & \nodetect & \pre \\ \cline{2-9}

 & Hacking in Darkness~\cite{lee2017hacking}
 & \detect & \detect & \detect & \pre
 & \parpre & \parpre & \pre \\ \cline{2-9}

 & Guards Dilemma~\cite{DBLP:conf/uss/BiondoCDFS18}
 & \detect & \detect & \detect & \pre
 & \parpre & \parpre & \pre \\ \cline{2-9}

 & COIN~\cite{khandaker2020coin}
 & \nodetect & \detect & \detect & \nodetect
 & \nodetect & \nodetect & \pre \\ \cline{2-9}

 & CVE-2017-6296~\cite{CVE-2017-6296}
 & \nodetect & \detect & \detect & \nodetect
 & \nodetect & \nodetect & \pre \\ \cline{2-9}

 & AsynShock~\cite{weichbrodt2016asyncshock}
 & \nodetect & \detect & \detect & \nodetect
 & \nodetect & \nodetect & \pre \\ \cline{2-9}

 & SmashEx~\cite{cui2021smashex}
 & \detect & \detect & \detect & \pre
 & \nodetect & \nodetect & \pre \\ \hline

AV2 & SGX DUMP~\cite{yoon2022sgxdump}
 & \detect & \detect & \nodetect & \nodetect
 & \parpre & \parpre & \pre \\ \hline

AV3 & M1racles~\cite{m1racles2021}
 & \nodetect & \nodetect & \nodetect & \nodetect
 & \nodetect & \nodetect & \pre \\ \hline

AV4 & CVE-2016-10423~\cite{cve2016_10423}
 & \nodetect & \nodetect & \nodetect & \nodetect
 & \parpre & \nodetect & \pre \\ \hline
\end{tabular}}
\label{tab:related-work}
\end{table}

We categorize data leakage attacks on enclaves in Table~\ref{tab:related-work} into four Attack Vector classes: {\bf AV1:} vulnerabilities in enclave code or TEE design exploited to leak data directly via a {\tt memcpy} or an equivalent function, \textbf{AV2:} vulnerabilities in enclave code or design exploited to leak data indirectly via shared software structures, \textbf{AV3:} hardware flaws exploited to leak data via shared hardware registers, and \textbf{AV4:} hardware flaws exploited to leak data via shared I/O peripherals. Existing defenses against these attacks are discussed in Section~\ref{sec:rw}.
{\flushleft\textbf{AV1:}} Consider an attack that hijacks the control flow inside the enclave by exploiting a vulnerability. The attacker then chains together the gadgets to execute a sequence of {\tt load} and {\tt store} instructions that load sensitive data into a register and attempt to store it into non-enclave memory (Listing~\ref{lst:av1}(Top)). 
\begin{listing}[h]
  \centering

  \begin{minipage}[t]{0.45\textwidth}
    \begin{tcolorbox}[
        enhanced,
        colback=white, colframe=black, boxrule=0.6pt, sharp corners,
        left=6pt, right=1pt, top=0pt, bottom=0pt,
        title={Attacker-executed gadgets},
        colbacktitle=white, coltitle=black,
        boxed title style={empty, size=small},   
        attach boxed title to top center={yshift=-1.2mm}, 
      ]
\begin{lstlisting}[basicstyle=\ttfamily\small, frame=none, aboveskip=0pt, belowskip=0pt]
load r1, secret
store r1, non-enclave
\end{lstlisting}
    \end{tcolorbox}
  \end{minipage}
  \hfill
  \begin{minipage}[t]{0.45\textwidth}
    \begin{tcolorbox}[
        enhanced,
        colback=white, colframe=black, boxrule=0.6pt, sharp corners,
        left=6pt, right=1pt, top=0pt, bottom=0pt,
        title={\kingsguard\ enforced checks},
        colbacktitle=white, coltitle=black,
        boxed title style={empty, size=small},
        attach boxed title to top center={yshift=-1.2mm},
      ]
\begin{lstlisting}[basicstyle=\ttfamily\small, frame=none, mathescape=true, aboveskip=0pt, belowskip=0pt]
r1_t = 1 and addr == non-enclave
$H_{current} \neq H \in H^*$
r1 = 0
\end{lstlisting}
    \end{tcolorbox}
  \end{minipage}

  \caption{Preventing direct leakage to non-enclave memory.}
  \label{lst:av1}
\end{listing}

\kingsguard prevents such data leakage. The \texttt{secret} is tainted and when it is loaded into register {\tt r1}, the associated taint {\tt r1\_t} is set to 1 (Listing~\ref{lst:av1}(Bottom)). When the attacker tries to store the value in {\tt r1} to a non-enclave address, \kingsguard hardware calls the declassification module, which checks the current cumulative hash $H_{current}$ against pre-computed ADPs, $H^*$. Since the attacker has subverted execution, $H_{current}$ will not match any hash in $H^*$, and hence, the store will be prevented by overwriting the {\tt secret} in r1 to zeroes.
{\flushleft\textbf{AV2:}} We take as example the SGXDump~\cite{yoon2022sgxdump} attack. SGXDump does not directly copy data from the enclave to non-enclave memory, but encodes enclave data as addresses in the non-enclave memory and uses page table access bits to infer this data (Listing~\ref{lst:av2}).
\begin{listing}[h]
  \centering

  \begin{minipage}[t]{0.45\textwidth}
    \begin{tcolorbox}[
        enhanced,
        colback=white, colframe=black, boxrule=0.6pt, sharp corners,
        left=6pt, right=1pt, top=0pt, bottom=0pt,
        title={Attacker-executed gadgets},
        colbacktitle=white, coltitle=black,
        boxed title style={empty, size=small},   
        attach boxed title to top center={yshift=-1.2mm}, 
      ]
\begin{lstlisting}[basicstyle=\ttfamily\small, frame=none, aboveskip=0pt, belowskip=0pt]
load r1, secret
shift_left r1, 12
load r2, non-enclave array address
add r1, r1, r2
load r3, (r1)
\end{lstlisting}
    \end{tcolorbox}
  \end{minipage}
  \hfill
  \begin{minipage}[t]{0.45\textwidth}
    \begin{tcolorbox}[
        enhanced,
        colback=white, colframe=black, boxrule=0.6pt, sharp corners,
        left=6pt, right=1pt, top=0pt, bottom=0pt,
        title={\kingsguard\ enforced checks},
        colbacktitle=white, coltitle=black,
        boxed title style={empty, size=small},
        attach boxed title to top center={yshift=-1.2mm},
      ]
\begin{lstlisting}[basicstyle=\ttfamily\small, frame=none, mathescape=true, aboveskip=0pt, belowskip=0pt]
r1_t = 1 
r2_t  = 0
r1_t = r1_t $\lor$ r2_t = 1
r1_t = 1 and r1 == non-enclave addr
r1 = $a_{fixed}$
\end{lstlisting}
    \end{tcolorbox}
  \end{minipage}

  \caption{Preventing indirect leakage to non-enclave memory.}
  \label{lst:av2}
\end{listing}

In this example, the attacker hijacks the control flow inside the enclave to execute the sequence of instructions shown in Listing~\ref{lst:av2}(Top). The attacker loads a sensitive value into a register, converts this value into a page number (left shifting the register by 12 bits), accesses a non-enclave address in the computed page number, which can be detected by reading the access bit of the corresponding page in the page table entry. With this information, the attacker can transmit the value of {\tt secret} outside the enclave. \kingsguard prevents such data leakage by blocking access to tainted non-enclave addresses from the enclave code. Similar to AV1, the taint {\tt r1\_t} is set to 1 when {\tt secret} is loaded (Listing~\ref{lst:av2}(Bottom)). {\tt r2} is not tainted because it loads a non-enclave address. However, the final address is computed by adding the page number in {\tt r1} and the non-enclave address in {\tt r2}. The final address computed in {\tt r1} is also tainted due to the initial taint of {\tt r1}. \kingsguard hardware detects access to non-enclave memory and the taint associated with address register {$\tt r1\_t$}, and replaces the address in {\tt r1} with a fixed valid address in non-enclave memory, $a_{fixed}$.

{\noindent\textbf{AV3:}} For this attack vector, we consider a shared hardware register that can be misused to leak enclave data (Listing~\ref{lst:av3}).
\begin{listing}[h]
  \centering

  \begin{minipage}[t]{0.45\textwidth}
    \begin{tcolorbox}[
        enhanced,
        colback=white, colframe=black, boxrule=0.6pt, sharp corners,
        left=6pt, right=1pt, top=0pt, bottom=0pt, 
        title={Attacker-executed gadgets},
        colbacktitle=white, coltitle=black,
        boxed title style={empty, size=small},   
        attach boxed title to top center={yshift=-1.2mm}, 
      ]
\begin{lstlisting}[basicstyle=\ttfamily\small, frame=none, aboveskip=0pt, belowskip=0pt]
load r1, secret   
move user_config_reg, r1
context-switch 
move r1, user_config_reg  
\end{lstlisting}
    \end{tcolorbox}
  \end{minipage}
  \hfill
  \begin{minipage}[t]{0.45\textwidth}
    \begin{tcolorbox}[
        enhanced,
        colback=white, colframe=black, boxrule=0.6pt, sharp corners,
        left=6pt, right=1pt, top=0pt, bottom=0pt,
        title={\kingsguard\ enforced checks},
        colbacktitle=white, coltitle=black,
        boxed title style={empty, size=small},
        attach boxed title to top center={yshift=-1.2mm},
      ]
\begin{lstlisting}[basicstyle=\ttfamily\small, frame=none, mathescape=true, aboveskip=0pt, belowskip=0pt]
r1_t = 1
user_config_reg.owner = EID1
context-switch
EID1 $\neq$ currEID
user_config_reg = 0
\end{lstlisting}
    \end{tcolorbox}
  \end{minipage}
  \caption{Preventing indirect leakage via shared hardware registers.}
  \label{lst:av3}
\end{listing}
In this example, the attacker loads {\tt secret} into a register {\tt r1}. This taints {\tt r1}, setting {$\tt r1\_t$} to 1. The attacker then writes this tainted data to a shared register accessible in user space ({$\tt user\_config\_reg$}). \kingsguard detects this write of tainted data to the shared register and hence, stamps it with {\tt EID}. When the context switches outside the enclave, general-purpose registers like {\tt r1} are cleared. However, shared registers like configuration registers are not cleared and may retain their values. The non-enclave code or another malicious enclave may try to read the value from the shared register. \kingsguard detects the read from a shared register that has been stamped with a different {\tt EID}, and prevents the attacker from reading {\tt secret} by overwriting the register with zeroes (Listing~\ref{lst:av3}(Bottom)). 

{\noindent\textbf{AV4:}}
Data leakage in TEEs may also occur through shared peripherals. For example, a shared SPI bus in TrustZone allows one application to read data from another application's SPI connection~\cite{cve2016_10423}. A code snippet is shown in Listing~\ref{lst:av4}(Top). In \kingsguard, all device drivers operate outside the enclave, and all data written from the enclave to the peripherals must go through an ADP.

\begin{listing}[h]
  \centering

  \begin{minipage}[t]{0.46\textwidth}
    \begin{tcolorbox}[
        enhanced,
        width=\linewidth,
        colback=white, colframe=black, boxrule=0.6pt, sharp corners,
        left=6pt, right=1pt, top=0pt, bottom=0pt,
        title={Attacker-executed gadgets},
        colbacktitle=white, coltitle=black,
        boxed title style={empty, size=small},   
        attach boxed title to top center={yshift=-1.2mm}, 
      ]
\begin{lstlisting}[basicstyle=\ttfamily\small, frame=none, aboveskip=0pt, belowskip=0pt, breaklines=true, breakatwhitespace=true, columns=fullflexible]
load r1, secret     
store r1, SPI_BUF     //hijacked enclave code
context switch
read r2, SPI_BUF      //non-enclave code
\end{lstlisting}
    \end{tcolorbox}
  \end{minipage}
  \hfill
  \begin{minipage}[t]{0.46\textwidth}
    \begin{tcolorbox}[
        enhanced,
        width=\linewidth,
        colback=white, colframe=black, boxrule=0.6pt, sharp corners,
        left=6pt, right=1pt, top=0pt, bottom=0pt,
        title={\kingsguard\ enforced checks},
        colbacktitle=white, coltitle=black,
        boxed title style={empty, size=small},
        attach boxed title to top center={yshift=-1.2mm},
      ]
\begin{lstlisting}[basicstyle=\ttfamily\small, frame=none, mathescape=true, aboveskip=0pt, belowskip=0pt, breaklines=true, breakatwhitespace=true, columns=fullflexible]
r1_t = 1
SPI addr = non-enclave
$H_{current} \neq H \in H^*$
r1 = 0
\end{lstlisting}
    \end{tcolorbox}
  \end{minipage}
  \caption{Preventing data leakage via shared peripherals.}
  \label{lst:av4}
\end{listing}
Here, the attacker writes {\tt secret} to the memory-mapped SPI buffer. Since peripheral addresses are outside the enclave memory, any store to these addresses goes via declassification and will be prevented if unauthorized. \kingsguard assumes that DMA accesses to enclave memory are disallowed by the platform. This is consistent with the threat models and architectural assumptions adopted by prior TEE designs~\cite{DBLP:journals/iacr/sgxCostanD16, DBLP:conf/uss/sanctumCostanLD16, DBLP:conf/eurosys/keystoneLeeKSAS20}.

\begin{figure}[!t]
    \includegraphics[width=0.35\textwidth]{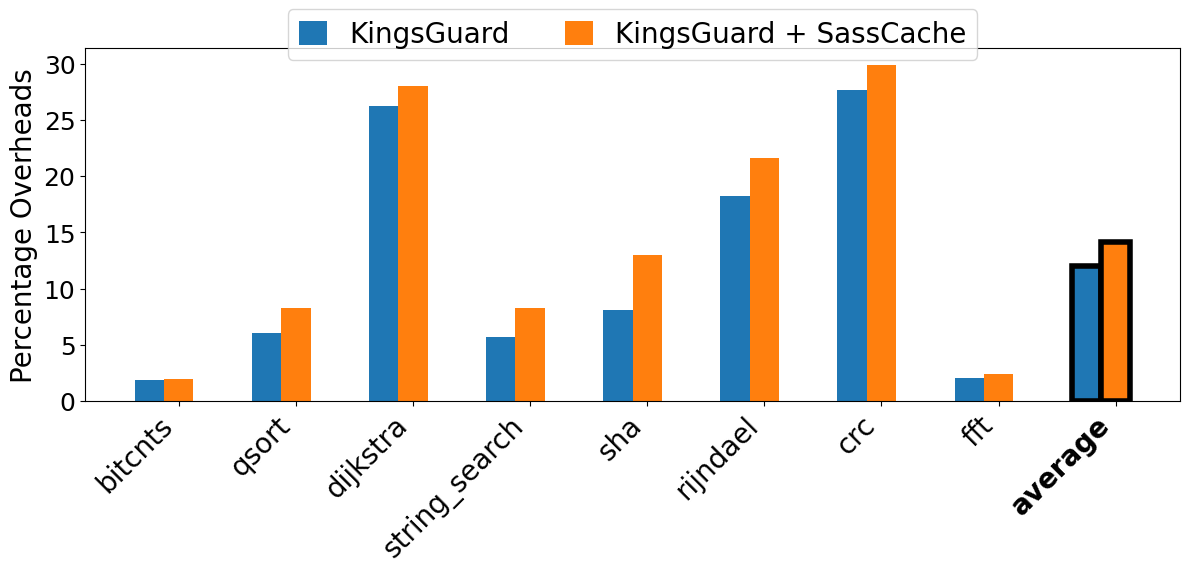}
     \caption{Percentage overheads of \kingsguard over the baseline processor with and without SassCache countermeasure.}
    \label{fig:sass}
\end{figure}

{\flushleft\textbf{Side Channel Attacks:} } TEEs are known to be vulnerable to side-channel attacks and \kingsguard is no exception. 
However, \kingsguard has been designed to be compatible with existing side-channel countermeasures like cache randomization~\cite{tan2020phantomcache, DBLP:conf/uss/WernerUG0GM19scattercache, DBLP:conf/micro/Qureshi18} and partitioning~\cite{DBLP:conf/uss/sanctumCostanLD16, DBLP:conf/micro/mi6, DBLP:conf/ndss/ArikanFCMWLA024}. To demonstrate that \kingsguard can be easily extended with existing side-channel defenses, we integrate SassCache~\cite{DBLP:conf/sp/GinerSPEUMG23} into our implementation. SassCache is a randomized cache with secure spacing that eliminates the attacker’s capability of building an eviction set in 99.99997\% of the cases. 
Figure~\ref{fig:sass} reports the performance overheads of \kingsguard when extended with SassCache. The results show that incorporating a cache-based side-channel countermeasure introduces a marginal overhead, amounting to an average of 1.93\%.

%% file: RelatedWork.tex
Existing defenses against TEE vulnerabilities are broadly classified as static or dynamic. Static techniques, such as symbolic execution~\cite{cloosters2020teerex, wang2023symgx}, fuzzing~\cite{cloosters2022sgxfuzz, khan2023fuzzing}, and taint analysis~\cite{stella}, identify the bugs before deployment but do not guarantee coverage, leaving enclaves exposed to attacks at runtime. 
Table~\ref{tab:related-work} focuses on runtime defenses that operate at either the hardware or software level.

{\noindent\textbf{Hardware Defenses.}} Hardware approaches~\cite{cerdeira2022rezone, Gu2022AHC} partition enclaves into fine-grained compartments to limit the impact of exploits. However, compromising a compartment that contains sensitive data directly exposes that data. In contrast, \kingsguard focuses on preventing sensitive data from leaving the enclave, rather than isolating regions of enclave code.

{\noindent\textbf{Software Defenses.}} Two common software-based runtime defenses involve verifying execution integrity~\cite{morbitzer2023guarantee,DBLP:conf/acsac/ToffaliniP0C22} and Dynamic Information Flow Control (DIFC)~\cite{tarkhani2023deluminator, DBLP:journals/corr/abs-2401-08901, DBLP:conf/uss/TsaiSJMPP20}. 
Both these approaches have considerable performance costs, exceeding 200\%~\cite{morbitzer2023guarantee, DBLP:conf/uss/TsaiSJMPP20}. Execution integrity is achieved via Control Flow Attestation~(CFA) in~\cite{morbitzer2023guarantee} and Provenace Analysis~(PA) in~\cite{DBLP:conf/acsac/ToffaliniP0C22}.
CFA verifies that an enclave’s control flow follows the expected execution path. By detecting deviations, CFA prevents data exfiltration caused by code reuse attacks. On the other hand, PA monitors instructions about the enclave state in addition to performing CFA, enabling it to detect TOCTOU attacks that preserve the expected control flow and therefore evade CFA. These defenses can detect attack vectors in AV1 and AV2 that rely on control flow hijacking (and TOCTOU in case of PA), but fail to address attack vectors AV3 and AV4. 

DIFC-based schemes enforce explicit information-flow policies to stop secrets from reaching untrusted sinks. In addition to the huge overheads, DIFC schemes also demand extensive changes to the OS or enclave code~\cite{tarkhani2023deluminator}. DIFC schemes can address the attack vector AV1 but fail to address AV2, AV3, and AV4. Compared to these solutions,  \kingsguard has an average performance overhead of only 5.69\%, requires minimal changes in the enclave code and OS, and is capable of addressing all four attack vectors (AV1 to AV4).

%% file: Conclusion.tex
\kingsguard addresses a critical blindspot in TEE designs---an assumption that enclave software and the TEE hardware are flawless. It introduces a hardware-centric defense that remains effective even when these assumptions fail. By integrating information flow tracking and declassification, \kingsguard is the first TEE that ensures that information only leaves the enclave through authorized paths and achieves this without OS trust. Besides simple annotations that are needed for sensitive variables and control paths, \kingsguard's design is transparent, requiring minimal changes in the application source code and OS. Our implementation of \kingsguard on a RISC-V processor shows an overhead of 10.8\% in hardware and 5.69\% in performance. 